\documentclass[letterpaper,pra,showpacs,superscriptaddress,twocolumn]{revtex4}
\usepackage{graphicx}%
\usepackage{bm}%
\usepackage%
{amsmath}%
\usepackage{amssymb}%
\newlength{\twocolumnwidth}
\newlength{\auxlv}
\begin{document}
\title{Quantum theory of an electromagnetic observer: classically behaving\ macroscopic systems and emergence of classical world in quantum electrodynamics}
\author{L.\ I.\ Plimak}
\affiliation{Institut f\"ur Quantenphysik, Universit\"at Ulm, 89069 Ulm, Germany}
\affiliation{Max Planck Institute for the Science of Light, 91058 Erlangen, Germany}
\affiliation{Max Born Institute for Nonlinear Optics and Short Pulse Spectroscopy, Division A1, 12489 Berlin, Germany}
\author{Misha Ivanov}
\affiliation{Max Born Institute for Nonlinear Optics and Short Pulse Spectroscopy, Division A1, 12489 Berlin, Germany}
\affiliation{Department of Physics, Imperial College London, South Kensington Campus, SW7 2AZ} 
\affiliation{Department of Physics, Humboldt University, Newtonstr.\ 15, 12489 Berlin, Germany} 
\author{A.\ Aiello}
\affiliation{Max Planck Institute for the Science of Light, 91058 Erlangen, Germany}
\affiliation{Institute for Optics, Information and Photonics, University of Erlangen-N\"urnberg, Staudtstrasse 7/B2, 91058 Erlangen, Germany}
\author{S.\ Stenholm}
\affiliation{Institut f\"ur Quantenphysik, Universit\"at Ulm, 89069 Ulm, Germany}
\affiliation{Physics Department, Royal Institute of Technology, KTH, Stockholm, Sweden}
\affiliation{Laboratory of Computational Engineering, HUT, Espoo, Finland}
\date{\today}
\begin{abstract}
Quantum electrodynamics\ under conditions of distinguishability of interacting matter entities, and of controlled actions and back-actions between them, is considered. Such ``mesoscopic quantum electrodynamics'' is shown to share its dynamical structure with the classical stochastic electrodynamics. In formal terms, we demonstrate that all general relations of the mesoscopic quantum electrodynamics\ may be recast in a form lacking Planck's constant. Mesoscopic quantum electrodynamics\ is therefore subject to ``doing quantum electrodynamics\ while thinking classically'', allowing one to substitute essentally classical considerations for quantum ones without any loss in generality. Implications of these results for the quantum measurement theory are discussed.

\end{abstract}
\maketitle
\section{Introduction}

Seeing a quantum effect, and even making practical use of it, does not necessarily imply recognising it as such. Magnetic compasses have been around for millenia, but quantum nature of ferromagnetism could not be understood till the discovery of the electron spin and the Pauli principle. More than forty years elapsed between Stewart's and Kirchhoff's experiments with thermal radiation and Planck's dissertation. Already in the 20th century, it took three years for Einstein to come up with a quantum explanation of Lenard's experiments. Clearly there must exist certain {\em mathematical patterns\/} in quantum equations of motion allowing one to look at the quantum without recognising it. 

An immediate reservation is in place here. This paper concerns, and is limited to, phase-space formulation of quantum electrodynamics based on the time-normal operator ordering \cite{MandelWolf,AgarwalWolf,VogelPF,PFunc,Itzykson,Schweber,Bogol,APII,DirResp,Maxwell}. This is only one example in the multitude of analogies existing between the quantum and classical mechanics (see, e.g., \cite{TwoDragomans}). Our work should be kept distinct from attempts to modify quantum electrodynamics\ by making arbitrary physically motivated assumptions about properties of matter (cf.\ \cite{Welton,BarutLamb}), as well as from attempts to imitate quantum behaviour in classical stochastic electrodynamics\ (cf.\ \cite{KaledStochED}). Another well established tradition is applying methods of quantum field theory\ to classical statistical mechanics \cite{Wyld,Nelson,ZinnJ,VasF,VasR}. 
Worth mentioning are also attempts to implement classical mechanics in the Hilbert space, see \cite{BondarPRL} and references therein. 

The result of this paper in a nutshell is that {\em quantum electrodynamics\ and classical stochastic electrodynamics\ share their macroscopic dynamical structure.\/} More specifically, all relations for generalised phase-space quasi-distributions \cite{MandelWolf,AgarwalWolf,VogelPF,PFunc} we obtain in this paper lack Planck's constant and, {\em ipso facto\/}, coincide with the corresponding relations of the classical stochastic electrodynamics\ for probability distributions. This lack of Planck's constant is exactly the aforementioned mathematical pattern. One way of formulating our results is that truly quantum dynamics is limited to microscopic conditions of indistinguishability and/or equations of motion of matter.

For purposes of this paper, {\em macroscopic\/} refers to electromagnetic interactions under conditions of distinguishability. 
Etymologies aside, we associate ``macroscopic'' and ``microscopic'' with ``distinguishable'' and ``indistinguishable'', and not with ``large'' and ``small''. 
``Distinguishability'' has its standard meaning (formalised as commutativity of relevant dynamical variables in the interaction picture). 
On the commonly used term {\em mesoscopic\/} see endnote \cite{endMeso}; we treat macroscopic and mesoscopic as synonyms.

The logic of the paper revolves around two concepts: ``obscured macroscopic view'' and ``doing quantum electrodynamics\ while thinking classically''. The former refers to a closed---albeit, strictly speaking, phenomenological---quantum dynamical framework confined to the electromagnetic\ field and current operators. 
The latter expresses the critical property of such framework: on rewriting it in the so-called {\em response picture\/} \cite{API,APII,APIII,WickCaus,Maxwell}, Planck's constant drops out. This is another manifestation of the said mathematical pattern. Any relation within ``obscured macroscopic view'' may therefore be obtained by formal quantisation of its classical counterpart. 

We stress that ``obscured macroscopic view'' is not an approximation, nor is it a modification of conventional quantum mechanics. It is a {\em voluntary limit\/} we impose on information extracted from a microscopic theory. It does not impose any restrictions on the theory itself, except that it be consistent with conventional quantum electrodynamics.

Implications of these results, with a multitude of specifications and reservations, are discussed at length in the paper. Here we allow ourselves only a brief comment so as to clarify our motivation. Particulars aside, quantum measurement is interaction of classical apparata with a quantum system. 
Nice as it may sound, this definition has two obvious problems. Classical apparata in the strict meaning of the term do not exist in nature, and it is not at all clear which laws---quantum or classical---should govern such interaction. 
Both problems disappear if we limit ``measurement'' to electromagnetic interaction\ of an observer---e.g., a human being---with the rest of the world. ``Obscured macroscopic view'' is a formal expression of limitations of such ``electromagnetic\ observer''. The difference between the classical and the quantum for such observer is {\em only\/} in whether currents he/she sees may or may not be {\em phenomenologically\/} interpteted as stochastic c-numbers. If such interpretation happens to be possible for some macroscopic quantum system, the latter becomes phenomenologically indistinguishable from a classical system. We term such systems classically behaving. 
``Classical apparata'' are {classically behaving quantum system}s, i.e., {\em inherently quantum\/} systems which {\em appear classical\/} to the observer. Their interaction with other quantum systems is governed by laws of quantum electrodynamics, which in this case are structurally identical with classical electrodynamics. If all devices involved behave classically, the whole situation reverts to classical electrodynamics. In particular, the classical world we perceive in everyday life is a collection of classically behaving quantum systems. As we demonstrate in this paper, these leading considerations are fully justified by the formal structure of quantum electrodynamics.

The goal of this paper is twofold. Firstly, we extend the earlier analyses \cite{Corresp,API,APII,APIII,WickCaus,DirResp,Maxwell,RelCaus,RelCausMadrid} to distinguishable matter subsystems. We show that the {\em formal\/} response characterisation of a solitary electromagnetic\ device \cite{APII,Maxwell} becomes its {\em physical\/} characterisation as a radiation scatterer were this device interacting electromagneticaly with other macroscopic devices. This proves overall physical consistency of the response viewpoint. 
Secondly, we establish the connection between our approach and the conventional phase-space techniques, making all results, the old as well as the new, intuitive. To this end we reformulate everything in terms of {\em conditional P-functionals\/} \cite{VogelPF,PFunc}; these quantities generalise both {\em conditional probability distributions\/} of classical stochastics and {\em quasiprobability distributions\/} of the conventional phase-space techniques, to arbitrary nonlinear non-Markovian bosonic quantum systems. 

Our analyses imply generalisation of the conventional phase-space techniques, firstly, beyond the resonance, or rotating wave, approximation (RWA), and, secondly, to quantum response problems. 
Coherent states of the harmonic oscillator, which traditionally serve as an entry point to phase-space approaches, were introduced by Schr\"odinger as early as 1926 \cite{SchrCohSt}. That quantum dynamics of free bosonic systems maps to classical dynamics irrespective of the quantum state was firstly noticed by Feynman in his review on path integrals \cite{Feynman}. This understanding was instrumental in developing quantum theory of photodetection by Glauber \cite{GlauberPhDet}. Glauber's theory was initially formulated for free electromagnetic fields, then extended to interacting fields by Kelley and Kleiner \cite{KelleyKleiner}. However, de Haan \cite{deHaan} and later Bykov and Tatarskii \cite{BykTat,Tat} pointed out that Kelley-Kleiner's results are limited to the RWA. Taking them outside the RWA leads to causality violations (we associate causality with retardation, cf.\ also endnote \cite{endCaus}). Lifting this restriction by amending the Glauber-Kelley-Kleiner\ definition is one of the key results of \mbox{Refs.\ \protect\cite{Corresp,API,APII,APIII,WickCaus,DirResp,Maxwell,RelCaus,RelCausMadrid}}. 

The critical generalisation is inclusion of response. It establishes a link to a host of powerful ideas, notably, to the real-time quantum field theory\ 
\cite{%
KuboIrrevI,%
MartinSchwinger,%
KadanoffBaym,%
UmezawaMatsumotoTachiki,%
KuboTodaHashitsumeII,%
RammerSmith,%
LandsmanVanWeert,%
Bellac,%
KamenevLevchenko%
}, Schwinger-Perel-Keldysh's closed-time-loop\ formalism \cite{SchwingerC,Perel,Keldysh}, and, last but not least, to relativistic quantum field theory\ \cite{Itzykson,Schweber,Bogol}. (This link was a subject of \cite{DirResp}.) From a physical perspective, it allows one to make propagation of quantum signals in space-time the guiding principle of the whole investigation. In particular, propagation of the electromagnetic field\ underlies the aforementioned amendment of the concept of time-normal ordering. It also reveals the inherent link between operator noncommutativity and response \cite{APII,APIII}.

The pivotal formal point of these analyses is that the description of a quantum system in terms of time-normal averages of Heisenberg operators conditional on external sources turns out to be equivalent to the closed-time-loop\ framework. This kind of description was termed in \cite{API,APII,APIII} {\em response formulation\/} of a quantum system (also {\em response picture\/}, {\em response viewpoint\/}, or {\em response characterisation\/}). The said equivalence was proven in \cite{APII} for interacting bosons and in \cite{APIII} for interacting fermions (see also endnote \cite{endDerResp}). 

The closed-time-loop\ formulation is the most general formal framework known in quantum field theory. {\em Ipso facto\/}, our approach may be seen as an ultimate generalisation of the phase-space techniques in quantum electrodynamics\ (cf.\ endnote \cite{endUltGen}). Its ability to tackle relativistic problems with renormalisations was demonstrated in \cite{DirResp,Maxwell}. 

The way the paper is structured reflects our wish to make the results intuitive. ``Obscured macroscopic view'' as a formal viewpoint is mostly relegated to appendices. In the main body of the paper, we proceed by ``doing quantum electrodynamics\ while thinking classically''. We start from a brief overview of the results in \mbox{Sec.\ \ref{ch:85NC}}. In \mbox{Sec.\ \ref{ch:MA}}, we develop the ``classical yardstick'': a collection of general formulae, describing properties of devices and their interactions in classical stochastic electrodynamics. Quantum mechanics starts in \mbox{Sec.\ \ref{ch:51BL}}, which summarises formal definitions. Conditional P-functionals are introduced in \mbox{Sec.\ \ref{ch:NP}}. The conjecture that classical formulae of \mbox{Sec.\ \ref{ch:MA}} are in fact {\em exact\/} relations for P-functionals is verified in \mbox{Secs \ref{ch:B}} and \ref{ch:91RJ}. In \mbox{Sec.\ \ref{ch:S}}, we review the formal approach, concentrating on the interplay of the formal viewpoint and approximations. The results are discussed in \mbox{Sec.\ \ref{ch:Dis}}. In Appendix \ref{ch:Reg} we briefly touch upon the role of causality and regularisations in the classical electromagnetic\ self-action problem. 
Appendix \ref{ch:QR} summarises the results ``imported'' from earlier papers. In Appendix \ref{ch:LM}, we outline redefinitions allowing one to exclude passive linear devices (mirrors etc.) from explicit consideration.
Appendices \ref{ch:2DC} and \ref{ch:PhDet} provide formal support to ``pictorial derivations'' in \mbox{Sec.\ \ref{ch:S}}.

\begin{figure}[b]
\includegraphics[scale=0.4]{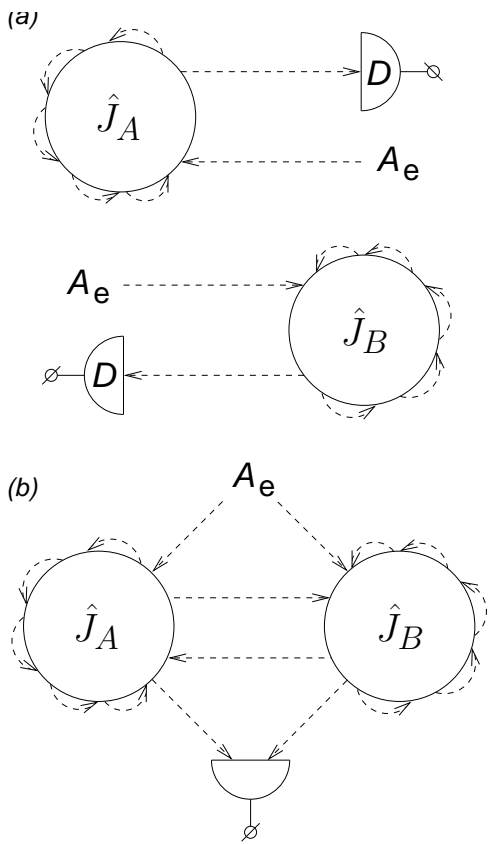}
\caption{From response characterisation of devices (a) to that of an interacting pair (b). Circles surrounded by curved dashed arrows symbolise ``dressed'' macroscopic devices, with all electromagnetic\ self-actions accounted for. Straight dashed arrows symbolise interactions of dressed devices. The detectors are a reminder that the time-normal ordering originates in photodetection theory. In the classical case of \mbox{Sec.\ \ref{ch:MA}}, disregard hats in the pictures and the remark on the time-normal ordering\ here.}
\label{fig:BirdsEye14}\end{figure}
\section{Overview}
\label{ch:85NC}%
In this paper we concern ourselves with the {\em formal structure\/} of quantum electrodynamics. Rather than attempting to calculate something measurable, we look at restrictions measurable quantities must obey, assuming the only means of accessing the world is the electromagnetic interaction. Moreover, we are only interested in relations which are independent of any particulars of nonelectromagnetic\ nature. This greatly limits the kind of question we are able to ask. 

Consider for instance electromagnetic\ interaction of two devices depicted schematically in \mbox{Fig.\ \ref{fig:BirdsEye14}}. For the time being, ignore hats and think about the devices as classical electromagnetic\ scatterers. In any real problem, there must exist microscopic dynamical models allowing one to calculate properties of the devices. 
These models necessarily rely on information about properties of matter, which is by itself nonelectromagnetic\ (e.g., the Dirac equation in spinor QED, or various models in solid state theory). 
Within ``obscured macroscopic view'', we are not inquiring about such details. We assume that properties of individual devices are known (\mbox{Fig.\ \ref{fig:BirdsEye14}}a), and ask how they should be formally combined to determine properties of the pair (\mbox{Fig.\ \ref{fig:BirdsEye14}}b).

This way, our investigation is to a large extent a highly formalised electronic-engineering viewpoint. When planning an experiment, an engineer is only interested in properties of a photodiod (say) such as frequency range, efficiency, dark count, dead time, saturation limit, and so on. From our perspective, all these are {\em response properties\/} of the photodiod. We mention features such as dark count and saturation limit to emphasise that proper formalisation of the engineering viewpoint cannot be too simple. It should include, e.g., such nastiness as {\em nonlinear non-Markovian self-noise\/}. This is the minimal level of theoretical sophistication accommodating for shot noise, dark count, dead time, and saturation. 

At the same time, the problem we face is largely pedagogical. We need a formal framework which would not impose any restrictions on models of devices. This implies a relatively high level of abstraction. Physically, all we can say about the arrangements in \mbox{Figs.\ \ref{fig:BirdsEye14}}a,b is that, 
\begin{itemize}
\item[1.]
in ``obscured macroscopic view'', each device is characterised by a c-number random current;
\item[2.]
statistics of this current is dependent (conditional) on the incoming field radiated by some external sources and affecting the device;
\item[3.]
when characterising a particular device, the incoming field should be regarded given; 
\item[4.]
for each device in the pair, the incoming field is the external field \mbox{$A_{\mathrm{e}}$} plus radiation of the other device;
\item[5.]
the current characterising the pair is a sum of currents in the devices;
\item[6.]
the field emitted by the pair is a sum of fields emitted by the devices.
\end{itemize}
In \mbox{Sec.\ \ref{ch:MA}}, we take these trivialities in classical electrodynamics\ to a high level of formal abstraction. In the following sections, we demonstrate that quantum-electrodynamical relations in the ``obscured macroscopic view'' express nothing but the same set of trivialities. It is this fact---{\em and not the trivialities themselves\/}---that constitutes the result of this paper. 

\section{``Obscured macroscopic view'' in classical electrodynamics}%
\label{ch:MA}%
\subsection{Preliminary remarks%
\label{ch:51BV}}

In this section we construct a ``classical yardstick'' for future quantum analyses. 
We develop a formal response characterisation of a device in classical stochastic electrodynamics, including the formal solution to the self-action problem (cf.\ \mbox{Fig.\ \ref{fig:BirdsEye14}}a; ignore hats), then derive formulae reducing properties of an interacting pair to its components (cf.\ \mbox{Fig.\ \ref{fig:BirdsEye14}}b).
We employ an intuitive characterisation of radiation scatterers in terms of probability distributions conditional on external sources, making all our derivations next to trivial. 

\subsection{Macroscopic device in classical stochastic electrodynamics\label{ch:34TD}}
\subsubsection{Conditional probability functional%
\label{ch:39TK}}
From the point of view of an ``electromagnetic observer'', the world is a maze of electromagnetic\ currents. 
These currents are mostly stochastic (we do not need quantum mechanics\ to arrive at this conclusion, experimental evidence is aplenty). Thinking of a device as a macroscopic radiation scatterer (\mbox{Fig.\ \ref{fig:BirdsEye14}}a), we characterise it by a probability distribution \mbox{$p\protect{ [ \mathcal{J}| A_{\mathrm{e}} ]}$} over a random current \mbox{$\mathcal{J}(t)$} conditional on the incoming (external) field \mbox{$A_{\mathrm{e}}(t)$}. For simplicity, we suppress all arguments of the fields and currents except time; for a generalisation beyond single mode see \mbox{Sec.\ \ref{ch:53LT}} below. 

In a general nonlinear non-Markovian case, \mbox{$p\protect{ [ \mathcal{J}| A_{\mathrm{e}} ]}$} is a nontrivial {\em functional\/} of two variables. Its full name is {\em conditional functional probability distribution.\/} In classical stochastic electrodynamics, 
\begin{align} p
\protect{ [ \mathcal{J}| A_{\mathrm{e}} ]}\geq 0, 
\label{eq:35AT}\end{align}%
but we never use this in our reasoning---otherwise ``doing quantum electrodynamics\ while thinking classically'' would certainly fail.

\subsubsection{Beyond the single mode%
\label{ch:53LT}}
Viewed literally, all formulae in this paper are written for a one-mode case. To adapt them to more complicated situations it suffices to ``expand'' the time variable. A multi-mode case is recovered replacing, 
\begin{align} 
\begin{aligned} 
 & t\to t,k, \quad \int dt \to \sum_k\int dt , \\ & \mathcal{J}(t) \to \mathcal{J}_k(t),\quad A_{\mathrm{e}}(t) \to A_{\mathrm{e}k}(t),\quad \cdots,
\end{aligned} 
\label{eq:54LU}\end{align}%
where $k$ is the mode index. (Here and hereafter, omitted integration limits indicate the maximal possible area of integration.) A nonrelativistic 3D formulation emerges on replacing, 
\begin{align} 
\begin{aligned} 
 & t\to t,{\mbox{\rm\boldmath$r$}},k, \quad \int dt \to \sum_k\int d^3{\mbox{\rm\boldmath$r$}}dt , 
\\ & \mathcal{J}(t) \to {\mbox{\rm\boldmath$\mathcal{J}$}}({\mbox{\rm\boldmath$r$}},t),\quad A_{\mathrm{e}}({\mbox{\rm\boldmath$r$}},t) \to {\mbox{\rm\boldmath$A$}}_{\mathrm{e}}({\mbox{\rm\boldmath$r$}},t),\quad \cdots,
\end{aligned} 
\label{eq:55LV}\end{align}%
where \mbox{$k=1,2,3$} is now the 3D index. This covers all cases of interest in macroscopic optics. Relativistic reformulation takes changing a few signs and factors, see \cite{Maxwell} and Appendix \ref{ch:GZ}.

\subsubsection{Stochastic averages as path integrals%
\label{ch:40TL}}
Averages of the random current \mbox{$\mathcal{J}(t)$} may be written as functional (path) integrals, 
\begin{align} 
\begin{aligned} 
\protect \langle \mathcal{J}(t) \rangle & = \int \mathcal{D}[\mathcal{J}]\, p\protect{ [ \mathcal{J}| A_{\mathrm{e}} ]}\mathcal{J}(t) , 
\\ 
\protect \langle \mathcal{J}(t)\mathcal{J}(t') \rangle & = \int \mathcal{D}[\mathcal{J}]\, p
\protect{ [ \mathcal{J}| A_{\mathrm{e}} ]}\mathcal{J}(t) \mathcal{J}(t'), 
\end{aligned} 
\label{eq:27AK}\end{align}%
etc. Averages (\ref{eq:27AK}) are by definition conditional on the source. We stress that we have not invented anything new, but simply explained what the innocuous symbol of classical averaging actually means. 

We think about a path integral as a multidimensional integral in discretised time, 
\begin{align} 
\int \mathcal{D}[\mathcal{J}] = \prod_t 
 \bigg [ \int d\mathcal{J}(t) \bigg ] ,
\label{eq:28AL}\end{align}%
cf.\ the remark on omitted integration limits after \mbox{Eq.\ (\ref{eq:54LU})}. To generalise (\ref{eq:28AL}) beyond a single mode, use formulae of \mbox{Sec.\ \ref{ch:53LT}}. Such definition makes all manipulations straightforward. A mathematically sound limit of continuous time is outside the scope of this paper. 
\subsubsection{Causality%
\label{ch:37TH}}
The minimal requirement for any physical model is {\em nonrelativistic causality\/}. Namely, the external field \mbox{$A_{\mathrm{e}}(t)$} at time $t$ may only influence the current \mbox{$\mathcal{J}(t')$} for times \mbox{$t'>t$} (cf.\ endnote \cite{endCaus}). Varying \mbox{$A_{\mathrm{e}}(t)$} for times \mbox{$t>t'$} should have no effect on \mbox{$\mathcal{J}(t')$}. Formally, this is expressed by the zero variational (functional) derivative, 
\begin{align} 
\frac{\delta \mathcal{J}(t')}{\delta A_{\mathrm{e}}(t)} & = 0, 
\quad t>t'. 
\label{eq:86ND}\end{align}%
For the averages, 
\begin{align} 
\begin{aligned} 
\frac{\delta \protect \langle \mathcal{J}(t') \rangle }{\delta A_{\mathrm{e}}(t)} & = 0, 
\quad t>t', 
\\
\frac{\delta \protect \langle \mathcal{J}(t')\mathcal{J}(t'') \rangle }{\delta A_{\mathrm{e}}(t)} & = 0, 
\quad t>t',t'', 
\end{aligned} 
\label{eq:54NF}\end{align}%
and similarly for higher-order averages. 

It should not be missed that neither \mbox{Eq.\ (\ref{eq:86ND})} nor (\ref{eq:54NF}) assume setting \mbox{$A_{\mathrm{e}}(t)$} to zero after differentiation. The latter is characteristic of the definition of generalised susceptibilities, 
\begin{align} 
\mathcal{R}(t_1',\cdots,t_m'|t_1,\cdots,t_n) 
=\frac{\delta^n \protect \langle \mathcal{J}(t'_1)\cdots\mathcal{J}(t'_m) \rangle }{\delta A_{\mathrm{e}}(t_1)\cdots\delta A_{\mathrm{e}}(t_n)}\Big | _{A_{\mathrm{e}}=0} .
\label{eq:88NF}\end{align}%
From (\ref{eq:54NF}) we obtain, 
\begin{align} 
\mathcal{R}(t_1',\cdots,t_m'|t_1,\cdots,t_n) 
=0, \quad \min \{ t \} >\max \{ t' \} . 
\label{eq:89NH}\end{align}%
However, \mbox{Eqs.\ (\ref{eq:54NF})} afford a stronger result, 
\begin{align} 
\frac{\delta^n \protect \langle \mathcal{J}(t'_1)\cdots\mathcal{J}(t'_m) \rangle }{\delta A_{\mathrm{e}}(t_1)\cdots\delta A_{\mathrm{e}}(t_n)}=0, 
\quad \min \{ t \} >\max \{ t' \} . 
\label{eq:90NJ}\end{align}%
Unlike (\ref{eq:89NH}), this condition, as well as the underlying conditions (\ref{eq:86ND}) and (\ref{eq:54NF}), remain physically meaningful also in a vicinity of a phase transition. 
\subsubsection{``Passive linear medium'' {\em versus\/} ``devices''%
\label{ch:91NK}}
The scattered field is the field radiated by the random current, 
\begin{align} 
\mathcal{A}(t) = \int dt' G_{\text{R}} (t,t')\mathcal{J}(t'), 
\label{eq:95RN}\end{align}%
where \mbox{$G_{\text{R}} $} is the transfer function of the linear medium (or vacuum) in which the device is immersed. Response of the linear medium should be causal, 
\begin{align} 
G_{\text{R}} (t,t') = 0, \quad t<t' .
\label{eq:92NL}\end{align}%
We do not assume the linear medium to be stationary (nor homogeneous---subject to generalisations as per \mbox{Sec.\ \ref{ch:53LT}}). We however do assume it to be passive: its radiation (e.g., thermal) must be negligible. 

For simplicity, we assume that \mbox{$G_{\text{R}} $} is not affected by the presence of devices. This assumption is utterly impractical. In a laser theory (say), it means that the mirrors and the active medium are introduced on an equal basis. It is much more common to reassign linear susceptibilies of devices to the ``free'' field. All passive linear devices (mirrors, beam-splitters and {\em tutti quanti\/}) are then eliminated from explicit consideration, being accounted for in \mbox{$G_{\text{R}} $}. For purposes of this paper, practicality is the least of concerns, so we stick with simplicity. We return to this assumption in \mbox{Sec.\ \ref{ch:39MQ}}; a formal way of lifting it is outlined in Appendix \ref{ch:LM}.

Whether a particular physical entity is regarded ``linear medium'' or ``device'' depends on the degree of insight. For instance, in nonrelativistic theories, vacuum is ``linear medium''. In relativistic QED, vacuum is rather a ``nonlinear device'', of which the {\em low frequency limit of linear response\/} is separated from ``vacuum corrections''. This low frequency limit is that what one calls ``the free electromagnetic field\ in vacuum''. The whole procedure is highly nontrivial because of divergences and renormalisations. For details, see textbooks \cite{Schweber,Bogol,Itzykson} and papers \cite{DirResp,Maxwell}.

\subsubsection{Radiation laws for averages%
\label{ch:38TJ}}
Mixed averages of the field and current reduce to current averages, 
\begin{gather} 
\begin{gathered} 
\protect \langle \mathcal{A}(t) \rangle = \int dt' G_{\text{R}} (t,t')\protect \langle \mathcal{J}(t') \rangle , \\
\protect \langle {\mathcal{J}(t)\mathcal{A}(t')} \rangle = \int dt'' G_{\text{R}} (t',t'')%
\protect \langle {\mathcal{J}(t)\mathcal{J}(t'')} \rangle , 
\end{gathered} 
\label{eq:50BK}\end{gather}%
and similarly for higher-order averages. 
Formally, these averages imply existence of a joint conditional probability functional over the current and field, 
\begin{align} 
\protect \langle \mathcal{J}(t)\mathcal{A}(t') \rangle & = \int \mathcal{D}[\mathcal{J}] \mathcal{D}[ \mathcal{A}]\, p
\protect{ [ \mathcal{J},\mathcal{A}| A_{\mathrm{e}} ]}\mathcal{J}(t) \mathcal{A}(t'), 
\label{eq:29AM}\end{align}%
etc. The radiation law (\ref{eq:95RN}) reduces \mbox{$p
\protect{ [ \mathcal{J},\mathcal{A}| A_{\mathrm{e}} ]}$} to \mbox{$p
\protect{ [ \mathcal{J}| A_{\mathrm{e}} ]}$}, 
\begin{align} 
p\protect{ [ \mathcal{J},\mathcal{A}| A_{\mathrm{e}} ]}= \delta 
 [ \mathcal{A}-G_{\text{R}} \mathcal{J}] p\protect{ [ \mathcal{J}| A_{\mathrm{e}} ]}, 
\label{eq:30AN}\end{align}%
where we use a condensed notation, 
\begin{align} 
G_{\text{R}} g(t) = \int dt' G_{\text{R}} (t,t')g(t'). 
\label{eq:77NV}\end{align}%
The $\delta $-functional is defined as a multidimensional $\delta $-function, 
\begin{align} 
\delta
 [ f ] = \prod_t \delta (f(t)) . 
\label{eq:31AP}\end{align}%
This matches \mbox{Eq.\ (\ref{eq:28AL})}, making all algebraic manipulations straightforward.
\subsection{Electromagnetic self-action%
\label{ch:36AU}}
Distribution \mbox{$p\protect{ [ \mathcal{J}| A_{\mathrm{e}} ]}$} provides a macroscopic characterisation of a device in terms of statistics of a random current, conditional on radiation of {\em other devices\/}. In a microscopic model of a device, evolution of matter depends on the local (microscopic) field \mbox{$A_{\mathrm{loc}}(t)$} rather than on the external field \mbox{$A_{\mathrm{e}}(t)$}. 
The local field includes self-radiation of matter, i.e., 
\begin{align} 
A_{\mathrm{loc}}(t) = A_{\mathrm{e}}(t) + G_{\text{R}} \mathcal{J}(t), 
\label{eq:37AV}\end{align}%
where we used condensed notation (\ref{eq:77NV}). The microscopic model allows one to calculate statistics of the current conditional on the local field. Formally, it is described by the probability distribution \mbox{$p^{\mathrm{I}}\protect{ [ \mathcal{J}| A_{\mathrm{loc}} ]}$}, which we assume to be known (\mbox{$A_{\mathrm{loc}}$} here is a given quantity; \mbox{Eq.\ (\ref{eq:37AV})} is disregarded). The self-action problem is then solved by the ``stohastic dressing formula'', 
\begin{align} 
p\protect{ [ \mathcal{J}| A_{\mathrm{e}} ]}= 
p^{\mathrm{I}}\protect{ [ \mathcal{J}| A_{\mathrm{e}}+G_{\text{R}} \mathcal{J}]}. 
\label{eq:38AW}\end{align}%
The apparent simplicity of this relation is extremely deceptive. Its consistency is subject to causality and suppression of instantaneous self-action. For a brief discussion see \mbox{Appendix \ref{ch:Reg}}.

\subsection{Interaction of distinguishable devices%
\label{ch:26SV}}
Consider now a pair of devices, labeled $A$ and $B$ (cf.\ \mbox{Fig.\ \ref{fig:BirdsEye14}}b; again ignore hats). Each is characterised by a pair of distributions connected by \mbox{Eq.\ (\ref{eq:38AW})}, macroscopically by \mbox{$p_{A,B}[\mathcal{J}|A_{\mathrm{e}}]$}, and microscopically by \mbox{$p_{A,B}^{\mathrm{I}}[\mathcal{J}|A_{\mathrm{loc}}]$}. \mbox{Equation (\ref{eq:38AW})} extends to individual devices, 
\begin{align} 
p_{A,B}\protect{ [ \mathcal{J}| A_{\mathrm{e}} ]}= 
p_{A,B}^{\mathrm{I}}\protect{ [ \mathcal{J}| A_{\mathrm{e}}+G_{\text{R}} \mathcal{J}]}. 
\label{eq:77LH}\end{align}%
As a compound device, the pair is characterised by \mbox{$p[\mathcal{J}|A_{\mathrm{e}}]$} and \mbox{$p^{\mathrm{I}}[\mathcal{J}|A_{\mathrm{loc}}]$}. 

Now, what does it mean that the devices are distinguishable? Since indistinguishability in the true meaning of the term does not occur in classical mechanics, the only assumption we have to make is that their models do not contain correlated noise sources. In other words, the currents $\mathcal{J}_A$ and $\mathcal{J}_B$ should be statisticaly independent if considered conditional on the local field. In formal terms, the corresponding joint probability distribution should factorise, 
\begin{align} 
\begin{aligned} 
p^{\mathrm{I}} [ 
\mathcal{J}_A, \mathcal{J}_B | A_{\mathrm{loc}}
 ] = p^{\mathrm{I}}_{A} [ 
\mathcal{J}_A | A_{\mathrm{loc}}
 ] p^{\mathrm{I}}_{B} [ 
\mathcal{J}_B | A_{\mathrm{loc}}
 ] . 
\end{aligned} 
\label{eq:4KR}\end{align}%
This realised, solution to the interaction problem in terms of the microscopic distributions is trivial. 
The total random current characterising the composite device is a sum of the currents in devices, 
\begin{align} 
\begin{aligned} 
\mathcal{J}(t) = \mathcal{J}_A(t)+\mathcal{J}_B(t). 
\end{aligned} 
\label{eq:3KQ}\end{align}%
Integrating (\ref{eq:4KR}) over the redundant information we find the probability distribution over the full current, 
\begin{align} 
p^{\mathrm{I}} [ 
\mathcal{J}| A_{\mathrm{loc}}
 ] & = \int \mathcal{D}[\mathcal{J}_A] \mathcal{D}[\mathcal{J}_B]\delta [\mathcal{J}-\mathcal{J}_A-\mathcal{J}_B]
 \nonumber\\&\quad \times p^{\mathrm{I}}_A[\mathcal{J}_A | A_{\mathrm{loc}}]p^{\mathrm{I}}_B[\mathcal{J}_B | A_{\mathrm{loc}}] 
. 
\label{eq:5KS}\end{align}%
We refrain from integration over \mbox{$\mathcal{J}_A$} or \mbox{$\mathcal{J}_B$} to maintain symmetry of formulae. 

The expression for the macroscopic functional \mbox{$p\protect{ [ \mathcal{J}| A_{\mathrm{e}} ]}$} by \mbox{$p_A\protect{ [ \mathcal{J}_A | A_{\mathrm{e}} ]}$} and \mbox{$p_B\protect{ [ \mathcal{J}_B | A_{\mathrm{e}} ]}$} also follows with ease. Applying the stochastic dressing formula (\ref{eq:38AW}) to (\ref{eq:5KS}) we obtain, 
\begin{align} 
p [ 
\mathcal{J}| A_{\mathrm{e}}
 ] & = \int \mathcal{D}[\mathcal{J}_A] \mathcal{D}[\mathcal{J}_B]\delta [\mathcal{J}-\mathcal{J}_A-\mathcal{J}_B]
 \nonumber\\&\quad \times p^{\mathrm{I}}_A[\mathcal{J}_A | A_{\mathrm{e}}+G_{\text{R}} \mathcal{J}]p^{\mathrm{I}}_B[\mathcal{J}_B | A_{\mathrm{e}}+G_{\text{R}} \mathcal{J}] 
. 
\label{eq:6KT}\end{align}%
We then ``dress'' the components of the device by making use of \mbox{Eqs.\ (\ref{eq:77LH})}, 
\begin{align} 
\begin{aligned} 
p^{\mathrm{I}}_A[\mathcal{J}_A | A_{\mathrm{e}}+G_{\text{R}} \mathcal{J}] & = 
p^{\mathrm{I}}_A[\mathcal{J}_A | A_{\mathrm{e}}+G_{\text{R}} (\mathcal{J}_A+\mathcal{J}_B)]
 \\& = p_A[\mathcal{J}_A | A_{\mathrm{e}}+G_{\text{R}} \mathcal{J}_B]
 , \\
p^{\mathrm{I}}_B[\mathcal{J}_B | A_{\mathrm{e}}+G_{\text{R}} \mathcal{J}] & = 
p^{\mathrm{I}}_B[\mathcal{J}_B | A_{\mathrm{e}}+G_{\text{R}} (\mathcal{J}_A+\mathcal{J}_B)]
 \\& = p_B[\mathcal{J}_B | A_{\mathrm{e}}+G_{\text{R}} \mathcal{J}_A]
 . 
\end{aligned} 
\label{eq:7KU}\end{align}%
Finally, 
\begin{align} 
\begin{aligned} 
p [ 
\mathcal{J}| A_{\mathrm{e}}
 ] & = \int \mathcal{D}[\mathcal{J}_A] \mathcal{D}[\mathcal{J}_B]\delta [\mathcal{J}-\mathcal{J}_A-\mathcal{J}_B]
 \\&\quad \times p_A[\mathcal{J}_A | A_{\mathrm{e}}+G_{\text{R}} \mathcal{J}_B]p_B[\mathcal{J}_B | A_{\mathrm{e}}+G_{\text{R}} \mathcal{J}_A] 
. 
\end{aligned} 
\label{eq:8KV}\end{align}%
Apart from \mbox{Eq.\ (\ref{eq:3KQ})}, this formula expresses the simple fact that external field affecting either device is \mbox{$A_{\mathrm{e}}$} plus radiation of the other device (cf.\ the list of ``trivialities'' in \mbox{Sec.\ \ref{ch:85NC}}). 

\section{The formal background%
\label{ch:51BL}}%
\subsection{The overall framework%
\label{ch:68UR}}%
In this section we reiterate formal definitions, allowing us to embed our analyses in a conventional quantum framework. We start with a brief overview of the general formal viewpoint and terminology. 

{\em Distinguishable subsystems\/} in quantum mechanics\ ``inhabit'' orthogonal subspaces of the Hilbert space. As a mathematical requirement, this definition applies in the interaction picture. All models in this paper comprise the {\em electromagnetic-field\/} subsystem and one or more {\em matter\/} subsystems. For brevity, we call matter subsystems {\em devices\/}. The term {\em bare device\/} applies to characterisation of a matter subsystem in the interaction picture. A {\em (solitary) dressed device\/} refers to a model where one of the matter subsystems is made to interact with the field subsystem. {\em Interacting devices\/} emerge by coupling more than one device to the field. Direct nonelectromagnetic\ interaction of devices is not allowed. 

Dressed solitary and interacting devices are characterised by the corresponding Heisenberg\ field and current operators (which are interaction-dependent, unlike the interaction-picture operators). Of main interest to us is how properties of interacting devices are related to those of the solitary ones. In formal terms, we look at relations between Heisenberg\ operators defined according to different interaction Hamiltonians. This point ought to be well noted: {\em we compare models\/}, rather than work with a single model.

Majority of the formal concepts we rely upon are standard, or at least well known. To the standard ones belong quantisation of the electromagnetic\ field, the electromagnetic\ Hamiltonian as a sum of matter, field and interaction Hamiltonians, the interaction {\em vs\/} Heisenberg\ picture, free and Heisenberg\ operators, and the initial (Heisenberg) state of the system specified in remote past. The latter gives rise to quantum averaging. We do not use the {Schr\"odinger}\ picture. Implicit in our techniques are also the evolution operator and the $\mathcal{S}$-matrix (they underlie the results ``imported'' from \cite{Maxwell}; cf.\ Appendix \ref{ch:QR}). All these concepts are amply covered in textbooks \cite{Louisell,Loudon,CohTanAtomPhoton,Itzykson,Schweber,Bogol}. Among the well-known concepts are the time-normal ordering\ of operators, introduced by Kelley and Kleiner and Glauber in the context of photodetection theory \cite{GlauberPhDet,KelleyKleiner,GlauberTN,MandelWolf}, and placing c-number sources into Hamiltonians, which we borrow from Kubo \cite{KuboIrrevI,KuboTodaHashitsumeII}. It should not be missed that, unlike Kubo, we do not limit our analyses to the linear response.

``Proprietary'' formal concepts we use are the {\em response transformation\/} of quantum mechanics \cite{API,APII,APIII,WickCaus,DirResp,Maxwell}, and the redefinition of the conventional time-normal\ operator ordering this transformation prompts \cite{APII}. The latter takes care of causality violations \cite{deHaan,BykTat,Tat} plagueing the conventional definition; this assures strict causality in our approach \cite{APII,APIII,RelCaus,RelCausMadrid}. Key points of the ``proprietary'' concepts will be reiterated where required. For details see the quoted papers.
\subsection{The model%
\label{ch:82VF}}%
\subsubsection{The Hamitonian%
\label{ch:67UQ}}%
Throughout the paper we assume the standard electromagnetic\ Hamiltonian, 
\begin{align} 
\hat H(t) & = \hat H_{\mathrm{m}}(t) + \hat H_{\mathrm{f}}(t) + \hat H_{\text{I}} (t). 
\label{eq:37MN}\end{align}%
Hereinafter m, f and I stand for {\em matter\/}, {\em field\/} and {\em interaction\/}. 
Quantum averaging is defined with respect to the factorised Heisenberg\ $\rho $-matrix, 
\begin{align} 
\protect \langle \cdots \rangle = \text{Tr}\hat \rho (\cdots), 
\quad 
\rho = \hat \rho _{\mathrm{f}}\otimes\hat \rho _{\mathrm{m}}, 
\label{eq:40BJ}\end{align}%
where \mbox{$\hat \rho _{\mathrm{f}}$} and \mbox{$\hat \rho _{\mathrm{m}}$} describe the initial states of field and matter. 
The same symbol (angle brackets) is used for the classical and quantum averages; what we have in mind should be clear in the context.

For simplicity we write the electromagnetic\ interaction in a single-mode form, 
\begin{align} 
\hat H_{\text{I}} (t) & = - \hat J(t)
 [ \hat A(t)+A_{\mathrm{e}}(t) ] , 
\label{eq:36MM}\end{align}%
where \mbox{$\hat J(t)$} and \mbox{$\hat A(t)$} are the current and the electromagnetic\ field operators in the interaction picture, and \mbox{$A_{\mathrm{e}}(t)$} is the aforementioned Kubo-style source. 
This Hamiltonian may be understood literally, or as a shorthand notation according to \mbox{Sec.\ \ref{ch:53LT}}. The latter implies summation over all field labels except time. E.g., in the multimode case, 
\begin{align} 
\hat J(t)
 [ \hat A(t)+A_{\mathrm{e}}(t) ] = \sum_{k} \hat J_k(t) [ \hat A_k(t)+A_{\mathrm{e}k}(t) ] , 
\label{eq:61PX}\end{align}%
in the nonrelativistic 3D case,
\begin{align} 
 & \hat J(t)
 [ \hat A(t)+A_{\mathrm{e}}(t) ] \nonumber\\&\quad = \int d^3{\mbox{\rm\boldmath$r$}}\sum_{k=1}^3 \hat J_k({\mbox{\rm\boldmath$r$}},t) [ \hat A_k({\mbox{\rm\boldmath$r$}},t)+A_{\mathrm{e}k}({\mbox{\rm\boldmath$r$}},t) ] , 
\label{eq:18ZZ}\end{align}%
etc.

Irrespective of all other formal details, \mbox{$\hat J(t)$} and \mbox{$\hat A(t')$} commute for arbitrary $t$ and $t'$, and all their mixed quantum averages factorise (cf.\ the remarks on the ``overall viewpoint'' in \mbox{Sec.\ \ref{ch:68UR}} above). That this does not extend to the Heisenberg\ operators---denoted \mbox{$\protect{\hat{\mathcal J}}(t)$} and \mbox{$\protect{\hat{\mathcal A}}(t)$}---is due solely to the interaction. The Heisenberg operators and their quantum averages are by construction dependent (conditional) on the external source \mbox{$A_{\mathrm{e}}$}. 
It should not be overlooked that interaction (\ref{eq:36MM}) is nonresonant: \mbox{$\hat J$} and \mbox{$\hat A$} are Hermitian operators, not broken into their frequency-positive and frequency-negative\ parts.

We adhere to certain notational conventions. Italics are consistently used for interaction-picture operators; this concerns Hamiltonians (\ref{eq:37MN}) and (\ref{eq:36MM}), as well as all other Hamiltonians in the paper. Calligraphic letters are equally consistently used for Heisenberg\ operators. 
\subsubsection{Microscopic dynamics%
\label{ch:61BW}}%

Apart from the minimal assumptions of Hermiticity of all operators and positivity of \mbox{$\rho $}-matrices, the matter Hamiltonian $\hat H_{\mathrm{m}}(t)$, the current operator $\hat J(t)$ and the state of the device $\hat \rho_{\mathrm{m}}$ may be arbitrary. 
Consistency of the response formulation \cite{API,APII,APIII} hinges on two properties of \mbox{$\hat A(t)$}: that it is a free Hermitian bosonic operator, of which the two-point commutator \mbox{$[\hat A(t),\hat A(t')]$} is a c-number, and that its frequency-positive\ part annihilates vacuum. We assume that the field is in a vacuum state, 
\begin{align} 
\rho _{\mathrm{f}} = \left|0\right\rangle\left\langle 0\right|. 
\label{eq:42BL}\end{align}%
Physically, assumptions (\ref{eq:40BJ}) and (\ref{eq:42BL}) matter little, because the system is allowed infinite time to develop.

{\em In no way do we assume that a closed system of microscopic equations of motion may be written solely in terms of the mesoscopic variables \mbox{$\hat J(t)$} and \mbox{$\hat A(t)$}.\/} Some underlying dynamical variables must exist, bosonic or fermionic, by which all matter variables are expressed, and for which closed equations of motion may be written. However, apart from the assumption that it may in principle be consistenly formulated, microscopic dynamics plays no role in our analyses. Macroscopic dynamics is observed, but in general does not allow one to construct a closed theory. Microscopic dynamics specifies the system completely, but may only be inferred from what is observed (cf.\ however \mbox{endnote \cite{endLinBosMeso}}). 

It is worthy of stressing that {\em the possibility of separation of the mesoscopic from microscopic dynamics is not an assumption\/}. It is a structural property of conventional quantum field theory\ (specifically, of the Feynman-Dyson approach) \cite{Itzykson,Schweber,Bogol}, cf.\ \cite{Maxwell} and Appendix \ref{ch:QR} in this paper.

\subsubsection{Running electromagnetic\ waves in quantum electrodynamics\label{ch:39MQ}}%
All parallels between classical and quantum electrodynamics\ are ultimately rooted on the {\em wave quantisation relation\/} \cite{Corresp}. It introduces the concept of running electromagnetic\ wave in quantum electrodynamics: 
\begin{align} 
 [ \hat A(t),\hat A(t') ] = -i\hbar 
 [ G_{\text{R}} (t,t')-G_{\text{R}} (t',t) ] . 
\label{eq:40MR}\end{align}%
The linear response function (a.k.a.\ transfer function or retarded propagator) of the electromagnetic field\ is found replacing the quantum current in (\ref{eq:36MM}) by a c-number source \mbox{$J_{\mathrm{e}}(t)$} and considering the linear response of the free field: 
\begin{align} 
G_{\text{R}} (t,t') = \frac{\delta 
\protect \langle \hat A(t) \rangle }{\delta J_{\mathrm{e}}(t')} 
\Big|_{J_{\mathrm{e}}=0}
= \frac{i}{\hbar }\theta(t-t') [ \hat A(t),\hat A(t') ] . 
\label{eq:41MS}\end{align}%
The first formula here is the definition of \mbox{$G_{\text{R}} $} and the last one is Kubo's relation for it \cite{KuboIrrevI,KuboTodaHashitsumeII}; we dropped quantum averaging of the commutator which is a c-number anyway. The wave quantisation formula follows trivially by inverting Kubo's one. 
It is not associated with any approximation, but its implications are much more profound under conditions of distinguishability. 
Note that the causality condition (\ref{eq:92NL}) is inherent to Kubo's response. 

In the main body of the paper, we assume that \mbox{$G_{\text{R}} $} is not affected by the presence of devices. This assumption was discussed in \mbox{Sec.\ \ref{ch:91NK}}. Apart from simplicity, the reason we stick to it is that it cannot be lifted in a Hamilton formulation. The corresponding formal redefinitions are summarised in Appendix \ref{ch:LM}. (Caveat: this Appendix relies on definitions yet to be given in \mbox{Sec.\ \ref{ch:54BP}}.) 

\subsubsection{Distinguishable devices}%
\label{ch:29SY}%
A pair of interacting distinguishable devices are introduced postulating,
\begin{gather} 
\begin{gathered} 
\hat H_{\mathrm{m}}(t) = \hat H_A(t) + \hat H_B(t) , \\
\hat J(t) = \hat J_A(t) + \hat J_B(t) , \quad
\hat \rho _{\mathrm{m}} = \hat \rho _A \otimes \hat \rho _B . 
\end{gathered} 
\label{eq:46MX}\end{gather}%
\mbox{$\hat J_A(t)$}, \mbox{$\hat J_B(t')$} and \mbox{$\hat A(t'')$} commute for arbitrary $t$, $t'$ and $t''$, and all their mixed averages factorise (cf.\ the remarks on Hilbert-space factorisation in \mbox{Sec.\ \ref{ch:68UR}}). 
We shall also need models of solitary devices, 
\begin{align} 
\hat H_{\mathrm{m}}(t) = \hat H_A(t), \quad 
\hat J(t) = \hat J_A(t), \quad 
\hat \rho _{\mathrm{m}} = \hat \rho _A , 
\label{eq:47MY}\end{align}%
for device $A$, and 
\begin{align} 
\hat H_{\mathrm{m}}(t) = \hat H_B(t), \quad 
\hat J(t) = \hat J_B(t), \quad 
\hat \rho _{\mathrm{m}} = \hat \rho _B ,
\label{eq:48MZ}\end{align}%
for device $B$.

\subsubsection{``Obscured macroscopic view''%
\label{ch:83VH}}%
Majority of the above definitions are formal. They serve to embed our analyses in a conventional dynamical quantum approach, and to perform its response transformation (cf.\ \mbox{Sec.\ \ref{ch:68UR}}); with that their role ends. Assumptions ``making it'' into the discussion are as follows: 
\begin{itemize}
\item 
The field is fully characterised by its response properties, reflected by \mbox{$G_{\text{R}} $}. 
\item 
Matter is fully characterised by quantum averages of products of the bare current operator, 
\begin{gather} 
\begin{gathered} 
\protect \langle \hat J(t_1)\cdots\hat J(t_m) \rangle . 
\end{gathered} 
\label{eq:31TA}\end{gather}%
\item 
For interacting devices, averages (\ref{eq:31TA}) factorise into products of the corresponding averages for devices, e.g., 
\begin{align} 
\protect \langle \hat J(t)\hat J(t') \rangle & = 
\protect \langle [ \hat J_A(t) + \hat J_B(t) ] [ \hat J_A(t') + \hat J_B(t') ] \rangle \nonumber\\& = 
\protect \langle \hat J_A(t)\hat J_A(t') \rangle +\protect \langle \hat J_A(t) \rangle \protect \langle \hat J_B(t') \rangle \nonumber\\&\quad + 
 \{ A\leftrightarrow B \} . 
\label{eq:84VJ}\end{align}%
\end{itemize}
That these assumptions suffice is a formal expression of consistency of ``obscured macroscopic view'' in quantum electrodynamics. 
\subsection{The time-normal ordering\ of operators%
\label{ch:33TC}}%
\subsubsection{The time-normal ordering\ beyond the rotating wave approximation%
\label{ch:05NY}}%
The time-normal ordering\ was introduced by Kelley and Kleiner and Glauber in the context of quantum photodetection theory \cite{GlauberPhDet,KelleyKleiner,GlauberTN} (for a pedagogical discussion see, e.g., Mandel and Wolf \cite{MandelWolf}). There are well-known causality violations associated with taking the Glauber-Kelley-Kleiner\ definition beyond the rotating wave approximation (RWA) \cite{deHaan,BykTat,Tat}, so that the idea that some amendments are in order should come naturally to the reader. These amendments were suggested by L.P.\ and S.S.\ in \cite{APII}, see also \cite{APIII,RelCaus,RelCausMadrid,Maxwell}. 

As an example, consider the time-normally ordered product of two current operators, \mbox{${\mathcal T}{\mbox{\rm\boldmath$:$}}\protect{\hat{\mathcal J}}(t)\protect{\hat{\mathcal J}}(t'){\mbox{\rm\boldmath$:$}}$}. The notation \mbox{${\mathcal T}{\mbox{\rm\boldmath$:$}}\cdots{\mbox{\rm\boldmath$:$}}$} for the time-normal ordering\ is borrowed from Mandel and Wolf \cite{MandelWolf}. 

We assume that the reader is familiar with the concept of separation of the frequency-positive and frequency-negative\ parts of functions. The latter is conveniently expressed as an integral transformation, (with \mbox{$f(t)$} being an arbitrary function)
\begin{gather} 
\begin{gathered} 
f^{(\pm)}(t) = \mathcal{F}^{(\pm)}_t f(t) = \int dt' \delta ^{(\pm)}(t-t') f(t'),\\ \delta ^{(\pm)}(t-t') = \pm\frac{1}{2\pi i(t-t'\mp i0^+)}.
\end{gathered} 
\label{eq:07PA}\end{gather}%
For more on this operation see, e.g., Appendix A in \cite{APII}. 

The beyond-the-RWA expression for \mbox{${\mathcal T}{\mbox{\rm\boldmath$:$}}\protect{\hat{\mathcal J}}(t)\protect{\hat{\mathcal J}}(t'){\mbox{\rm\boldmath$:$}}$} follows in two steps. Firstly, expand operators into the frequency-positive and frequency-negative\ parts, 
\begin{align} 
\protect{\hat{\mathcal J}}(t)=\protect{\hat{\mathcal J}}^{(+)}(t)+\protect{\hat{\mathcal J}}^{(-)}(t),
\label{eq:93RL}\end{align}%
and apply the Kelley-Kleiner\ definition, 
\begin{align} 
 & {\mathcal T}{\mbox{\rm\boldmath$:$}}\protect{\hat{\mathcal J}}(t)\protect{\hat{\mathcal J}}(t'){\mbox{\rm\boldmath$:$}}|_{\mathrm{KK}} \nonumber\\&\quad = 
{T_+\protect{\hat{\mathcal J}}^{(+)}(t)\protect{\hat{\mathcal J}}^{(+)}(t')} + 
{\protect{\hat{\mathcal J}}^{(-)}(t)\protect{\hat{\mathcal J}}^{(+)}(t')} \nonumber\\&\qquad + 
{\protect{\hat{\mathcal J}}^{(-)}(t')\protect{\hat{\mathcal J}}^{(+)}(t)} + 
{T_-\protect{\hat{\mathcal J}}^{(-)}(t)\protect{\hat{\mathcal J}}^{(-)}(t')} 
. 
\label{eq:94RM}\end{align}%
In this formula, \mbox{$T_{\pm}$} stand for the forward and reverse time orderings (often denoted by $T$ and $\tilde T$). Secondly, change the order of time orderings and the \mbox{$\mathcal{F}^{(\pm)}$} operations, 
\begin{align} 
 & {\mathcal T}{\mbox{\rm\boldmath$:$}}\protect{\hat{\mathcal J}}(t)\protect{\hat{\mathcal J}}(t'){\mbox{\rm\boldmath$:$}} \nonumber\\&\quad = 
\mathcal{F}_{t}^{(+)}\mathcal{F}_{t'}^{(+)}{T_+\protect{\hat{\mathcal J}}(t)\protect{\hat{\mathcal J}}(t')} + 
\mathcal{F}_{t}^{(-)}\mathcal{F}_{t'}^{(+)}{\protect{\hat{\mathcal J}}(t)\protect{\hat{\mathcal J}}(t')} \nonumber\\&\qquad + 
\mathcal{F}_{t'}^{(-)}\mathcal{F}_{t}^{(+)}{\protect{\hat{\mathcal J}}(t')\protect{\hat{\mathcal J}}(t)} + 
\mathcal{F}_{t}^{(-)}\mathcal{F}_{t'}^{(-)}{T_-\protect{\hat{\mathcal J}}(t)\protect{\hat{\mathcal J}}(t')} 
. 
\label{eq:19SN}\end{align}%
Time-normal products of many bosonic operators follow similarly. Note that 
\begin{align} 
{\mathcal T}{\mbox{\rm\boldmath$:$}}\protect{\hat{\mathcal J}}(t){\mbox{\rm\boldmath$:$}}= \protect{\hat{\mathcal J}}(t).
\label{eq:78LJ}\end{align}%

In (\ref{eq:19SN}), we moved the \mbox{$\mathcal{F}^{(\pm)}$} operations from under the averaging also in the second and third terms, where this does not matter. This serves to emphasise that the rhs of (\ref{eq:19SN}) is expressed by ``entire'' operators, not broken in their frequency-positive and frequency-negative\ parts 
(cf.\ Sec.\ 3.2 in \cite{Maxwell}). The reader familiar with the Schwinger-Perel-Keldysh closed-time-loop approach of quantum field theory\ \cite{SchwingerC,Perel,Keldysh,KamenevLevchenko} should have recoginised operator structures typical for this framework. Indeed, formally, \mbox{Eq.\ (\ref{eq:19SN})} and the like follow if using the wave quantisation formula to replace the matrix propagator in the closed-time-loop approach by the retarded propagator \cite{API,APII,APIII,WickCaus,DirResp,Maxwell}. This association with quantum field theory\ assures a solid formal ground for our analyses.
\subsubsection{Reality and causality%
\label{ch:06NZ}}%
It is straightforward to show that the time-normally ordered product of Hermitian operators is Hermitian; quantum averages of time-normally ordered products of Hermitian operators (time-normal averages, for short) are real. Causality is a subtler property. The following ``no-peep-into-the-future theorem'' was proven in \cite{RelCaus}: {\em a time-normally ordered product depends on operators it comprises only for times not larger than its largest time argument\/}. For a relativistic extension of this theorem see \cite{RelCaus,RelCausMadrid}; cf.\ also endnote \cite{endCaus}. 
\subsection{Condensed notation}%
\label{ch:DN}%
In addition to \mbox{Eq.\ (\ref{eq:77NV})}, we introduce three more cases of condensed notation:
\begin{align} 
fg & = \int dt f(t) g(t) , 
\label{eq:3VS}\\
fG_{\text{R}} g & = \int dtdt' f(t)G_{\text{R}} (t,t') 
g(t') , 
\label{eq:76NU}%
\\ 
fG_{\text{R}} (t) & = \int dt' g(t')G_{\text{R}} (t',t) , 
\label{eq:78NW}\end{align}%
where $f(t)$ and $g(t)$ are c-number or q-number functions. The ``products'' $fg$ and $fG_{\text{R}} g$ denote scalars, while $G_{\text{R}} g$ and $fG_{\text{R}} $---functions. 

The effect of condensed notation is twofold: it diminishes the bulk of formulae and reduces them visually to a scalar case. The latter much simplifies all calculations (we borrowed this trick from Vasil'ev \cite{VasF}). 
\section{Conditional time-normal averages and conditional P-functionals}%
\label{ch:NP}%
\subsection{Conditional time-normal averages%
\label{ch:06RY}}%
In this section we introduce the formal concepts, allowing one, so to speak, to implement \mbox{Fig.\ \ref{fig:BirdsEye14}} in quantum electrodynamics. We also use this opportunity to acquaint the reader with the key results of papers \cite{API,APII,APIII,WickCaus,DirResp,Maxwell,RelCaus,RelCausMadrid}. 

The quantum counterpart of the classical averages (\ref{eq:27AK}) are time-normal averages of the Heisenberg\ current operator \mbox{$\protect{\hat{\mathcal J}}(t)$}, 
\begin{align} 
\protect \langle \mathcal{J}(t)\mathcal{J}(t') \rangle \leftrightarrow
\protect \langle {\mathcal T}{\mbox{\rm\boldmath$:$}}\protect{\hat{\mathcal J}}(t)\protect{\hat{\mathcal J}}(t'){\mbox{\rm\boldmath$:$}}\rangle , 
\label{eq:53NE}\end{align}%
etc. The classical averages were dependent (conditional) on an external field \mbox{$A_{\mathrm{e}}(t)$} by definition. The quantum average exhibits this dependence by construction, due to the external field in the interaction (\ref{eq:36MM}). 

\mbox{Equation (\ref{eq:53NE})} introduces {\em response formulation\/} \cite{API,APII,APIII} (also response picture, response viewpoint, or response characterisation) of an electromagnetic\ device. Specifications beyond the RWA aside, one borrows the concepts of the photodetection theory to characterise the {\em output\/} of a device, while using Kubo-style sources to define the {\em input\/} (cf.\ also endnote \cite{endKuboSchwinger}). 
As remarked in \mbox{Sec.\ \ref{ch:05NY}} above, the beyond-the-RWA time-normal ordering\ emerges in fact within the closed-time-loop formalism of quantum field theory\ \cite{SchwingerC,Perel,Keldysh}. Moreover, the response formulation was shown to be formally equivalent to this formalism \cite{APII,APIII}. For details see the quoted papers and Appendix \ref{ch:QR}.

\subsection{Consistency condition%
\label{ch:64UM}}%

An immediate word of caution is in place here. In our approach external sources are used as functional arguments. This implies that they are {\em arbitrary\/}. Informally speaking, we allow c-number sources to be placed inside atomic nuclei. This extent of control is certainly impossible with real light sources. 
{\em Arbitrary external sources in quantum mechanics\ are unphysical quantities.\/} 

This controversy is resolved by the so-called {\em consistency condition\/} \cite{APII,APIII}. For an electromagnetic\ system, it is expressed by \mbox{Eqs.\ (\ref{eq:29FL})} and (\ref{eq:38FV}) in Appendix \ref{ch:QR}. It stipulates that, 
\begin{itemize}
\item
Response properties of a system are formally expressed in terms of Heisenberg\ operators defined with zero sources (cf.\ Kubo's \mbox{Eq.\ (\ref{eq:41MS})}).
\item
These properties are ``encoded'' in commutators of these operators (again, cf.\ \mbox{Eq.\ (\ref{eq:41MS})}).
\end{itemize}
These facts hold with given sources. With arbitrary sources, the consistency condition also shows that, 
\begin{itemize}
\item
The information lost due to fixing the time-normal order of operators is exactly that re-entering through the dependence of the ordered operator products on the sources. 
\end{itemize}
All these facts warrant the formal consistency of the response formulation, hence the term. 

Since arbitrary sources are unphysical, response formulation should be regarded an {\em intuitive interpretation\/} of a conventional quantum approach not relying on sources (specifically, of the aforementioned closed-time-loop\ formalism). This interpretation enables ``doing quantum electrodynamics\ while thinking classically'', making it extremely useful; nonetheless it is only formal. The interplay of the formal and the physical in response formulation will be further discussed in \mbox{Sec.\ \ref{ch:S}}. 

Being of fundamental importance for introduction of the response formulation, the consistency condition is of little---if any---relevance to analyses within it. For all practical purposes only its existence matters.

\subsection{Causality%
\label{ch:08PB}}%

The critical property making the analogy (\ref{eq:53NE}) meaningful is {\em causality\/}.
The no-peep-into-the-future theorem mentioned in \mbox{Sec.\ \ref{ch:06NZ}} assures validity of the quantum counterparts of \mbox{Eqs.\ (\ref{eq:54NF})}, 
\begin{gather} 
\begin{gathered} 
\frac{\delta }{\delta A_{\mathrm{e}}(t)}\protect \langle {\protect{\hat{\mathcal J}}(t')} \rangle = 0, 
\quad t>t',t', 
\\ 
\frac{\delta }{\delta A_{\mathrm{e}}(t)}\protect \langle {\mathcal T}{\mbox{\rm\boldmath$:$}}\protect{\hat{\mathcal J}}(t')\protect{\hat{\mathcal J}}(t''){\mbox{\rm\boldmath$:$}}\rangle = 0, 
\quad t>t',t'', 
\end{gathered} 
\label{eq:55NH}\end{gather}%
etc. Apart from the said theorem, they rely on the quantum counterpart of \mbox{Eq.\ (\ref{eq:86ND})}, 
\begin{align} 
\frac{\delta \protect{\hat{\mathcal J}}(t')}{\delta A_{\mathrm{e}}(t)} = 0, \quad t>t' . 
\label{eq:79LK}\end{align}%
Conditions (\ref{eq:55NH}) may be verified without making any assumptions about the quantum matter \cite{APII,APIII}. In other words, they constitute in essence a kinematical property of quantum mechanics. Relativistic extention of (\ref{eq:55NH}) takes a minimal assumption of commutativity of operators at space-like intervals. This is one of Wightman's axioms of quantum field theory \cite{WightmanGarding}. For details see \cite{RelCaus,RelCausMadrid} and endnote \cite{endCaus}. 

\subsection{Conditional P-functionals%
\label{ch:42TN}}
Generalised to quantum mechanics, \mbox{Eqs.\ (\ref{eq:27AK})} define the {\em functional conditional quasiprobability distribition\/}, or {\em conditional P-functional\/}, \mbox{$p
\protect{ [ \mathcal{J}| A_{\mathrm{e}} ]}$}, 
\begin{align} 
\begin{aligned} 
\protect \langle \protect{\hat{\mathcal J}}(t) \rangle & = \int \mathcal{D}[\mathcal{J}]\, p\protect{ [ \mathcal{J}| A_{\mathrm{e}} ]}\mathcal{J}(t) , 
\\ 
\protect \langle {\mathcal T}{\mbox{\rm\boldmath$:$}}\protect{\hat{\mathcal J}}(t)\protect{\hat{\mathcal J}}(t'){\mbox{\rm\boldmath$:$}}\rangle & = \int \mathcal{D}[\mathcal{J}]\, p
\protect{ [ \mathcal{J}| A_{\mathrm{e}} ]}\mathcal{J}(t) \mathcal{J}(t'), 
\end{aligned} 
\label{eq:33AR}\end{align}%
etc. These quantities generalise both {\em conditional probability distributions\/} of classical stochastics and {\em quasiprobability distributions\/} of the conventional phase-space techniques to arbitrary nonlinear non-Markovian quantum systems. We do not notationally distinguish the quasidistributions from probability distributions, assuming that, in classical mechanics, 
they are positive, 
while in quantum mechanics\ may become nonpositive. 

The term P-functional was introduced by W.\ Vogel \cite{VogelPF} for a functional generalisation of the conventional P-distribution \cite{MandelWolf}. Our definition generalises Vogel's, firstly, beyond RWA, and, secondly, to response. 
\subsection{Characteristic functionals}%
\label{ch:54BP}%
In classical mechanics, the conditional probability distribution \mbox{$p[\mathcal{J}|A_{\mathrm{e}}]$} is a primary quantity, while averages are defined in its terms. In quantum mechanics, all dynamical relations are derived for averages, and the corresponding P-functionals are to be defined in terms of the latter. 

We follow the association between probability distributions and characteristic functionals. For example, for the classical \mbox{$p[\mathcal{J}|A_{\mathrm{e}}]$} we have, 
\begin{align} 
\protect \langle \exp
 ( i\zeta \mathcal{J}) \rangle = \int \mathcal{D}[\mathcal{J}]\, p\protect{ [ \mathcal{J}| A_{\mathrm{e}} ]}\exp(i\zeta \mathcal{J}) . 
\label{eq:43TP}\end{align}%
We use condensed notation (\ref{eq:3VS}).
The quantum counterpart of (\ref{eq:43TP}) reads, 
\begin{align} 
\Phi _{\mathrm{m}}
\protect{ [ \zeta | A_{\mathrm{e}} ]} & = 
\protect \langle {\mathcal T}{\mbox{\rm\boldmath$:$}}\exp
 ( i\zeta \protect{\hat{\mathcal J}}) {\mbox{\rm\boldmath$:$}}\rangle \label{eq:20SP}\\ & = \int \mathcal{D}[\mathcal{J}]\, p\protect{ [ \mathcal{J}| A_{\mathrm{e}} ]}\exp
 ( i\zeta \mathcal{J}) . 
\label{eq:41AZ}\end{align}%
Apart from defining the conditional P-functional, we have also introduced notation for the corresponding characteristic one. We also need a quantum definition of the joint quasiprobability distribution, 
\begin{align} 
\Phi
\protect{ [ \eta ,\zeta | A_{\mathrm{e}} ]} & = 
\protect \langle {\mathcal T}{\mbox{\rm\boldmath$:$}}\exp
 ( i\eta \protect{\hat{\mathcal A}}+i\zeta \protect{\hat{\mathcal J}}) {\mbox{\rm\boldmath$:$}}\rangle \label{eq:44TQ}\\ & = \int \mathcal{D}[\mathcal{A}]\mathcal{D}[\mathcal{J}] \, p[\mathcal{A},\mathcal{J}|A_{\mathrm{e}}]\exp
 ( i\eta \mathcal{A}+i\zeta \mathcal{J}) . 
\label{eq:45TR}\end{align}%
cf.\ the classical \mbox{Eq.\ (\ref{eq:29AM})}.

Now, how should we define \mbox{$p^{\mathrm{I}}[\mathcal{J}|A_{\mathrm{loc}}]$}? 
To start with, \mbox{$A_{\mathrm{loc}}$} here is a given quantity; \mbox{Eq.\ (\ref{eq:37AV})} should be disregarded. 
Fixing \mbox{$A_{\mathrm{loc}}$} means that self-radiation is somehow prevented from affecting the device. This is achieved by suppressing interaction of the device with the quantised field, i.e., replacing interaction (\ref{eq:36MM}) by, 
\begin{align} 
\hat H_{\text{I}} (t) = - \hat J(t)A_{\mathrm{loc}}(t) . 
\label{eq:47BF}\end{align}%
This way, 
\begin{align} 
\Phi_{\mathrm{m}}^{\mathrm{I}} [
\zeta | A_{\mathrm{loc}}
 ] & = 
\protect \langle {\mathcal T}{\mbox{\rm\boldmath$:$}}\exp
 ( i\zeta \hat J_{\mathrm{loc}} ) {\mbox{\rm\boldmath$:$}}\rangle , 
\label{eq:21SQ}\\ & = \int \mathcal{D}[\mathcal{J}]\, p^{\mathrm{I}}\protect{ [ \mathcal{J}| A_{\mathrm{loc}} ]}\exp
 ( i\zeta \mathcal{J}) . 
\label{eq:24ST}\end{align}%
where \mbox{$\hat J_{\mathrm{loc}}$} is the Heisenberg\ current operator with respect to (\ref{eq:47BF}).

An immediate reservation is in place here. Definitions (\ref{eq:20SP})--(\ref{eq:24ST}) are a rare occasion when the consistency condition (see \mbox{Sec.\ \ref{ch:64UM}}) must be at least acknowledged. As explained there, \mbox{Eqs.\ (\ref{eq:20SP})}--(\ref{eq:24ST}) constitute an intuitive interpretation of all functionals involved, while their rigorous definitions cannot rely on external sources. Such definitions may be found in Appendices \ref{ch:UPC} and \ref{ch:25SU}. For the time being, awareness of their existence suffices.

This reservation is especially important for \mbox{$p^{\mathrm{I}}[\mathcal{J}|A_{\mathrm{loc}}]$}. It is in fact a primary quantity, with everything else expressed by it. That it can be {\em directly\/} expressed by averages of \mbox{$\hat J(t)$} is critical for consistency of our approach, cf.\ \mbox{Eq.\ (\ref{eq:31TA})} and comments thereon.

\subsection{Proof of \mbox{Eq.\ (\ref{eq:30AN})}%
\label{ch:46TS}}
As an example of working with functionals we prove \mbox{Eq.\ (\ref{eq:30AN})}. Substituting it into (\ref{eq:45TR}) results in the following relation between \mbox{$\Phi $} and \mbox{$\Phi _{\mathrm{m}}$}, 
\begin{align} 
\Phi
\protect{ [ \eta ,\zeta | A_{\mathrm{e}} ]}= \Phi_{\mathrm{m}}\protect{ [ \eta G_{\text{R}} + \zeta | A_{\mathrm{e}} ]}. 
\label{eq:47TT}\end{align}%
We use notation (\ref{eq:78NW}).
\mbox{Equation (\ref{eq:47TT})} is a special case of the general formula (\ref{eq:53ZE}) in Appendix \ref{ch:UPD}, verified in \cite{Maxwell}. \mbox{Equation (\ref{eq:30AN})} has thus been proven as a quantum formula for P-functionals.

Differentiating (\ref{eq:47TT}) and using the chain rule for functional derivatives we obtain quantum counterparts of \mbox{Eqs.\ (\ref{eq:50BK})}, 
\begin{gather} 
\begin{gathered} 
\protect \langle \protect{\hat{\mathcal A}}(t) \rangle = \int d t G_{\text{R}} (t- t')\protect \langle \protect{\hat{\mathcal J}}( t') \rangle , \\
\protect \langle {\mathcal T}{\mbox{\rm\boldmath$:$}}\protect{\hat{\mathcal J}}(t)\protect{\hat{\mathcal A}}(t'){\mbox{\rm\boldmath$:$}}\rangle = \int d t'' G_{\text{R}} (t'- t'')%
\protect \langle {\mathcal T}{\mbox{\rm\boldmath$:$}}\protect{\hat{\mathcal J}}( t)\protect{\hat{\mathcal J}}(t''){\mbox{\rm\boldmath$:$}}\rangle , \\
\begin{aligned} 
 & \protect \langle {\mathcal T}{\mbox{\rm\boldmath$:$}}\protect{\hat{\mathcal A}}(t)\protect{\hat{\mathcal A}}(t'){\mbox{\rm\boldmath$:$}}\rangle \\&\quad = \int d t'' d t''' G_{\text{R}} (t- t'') G_{\text{R}} (t'- t''')
\protect \langle {\mathcal T}{\mbox{\rm\boldmath$:$}}\protect{\hat{\mathcal J}}( t'')\protect{\hat{\mathcal J}}( t'''){\mbox{\rm\boldmath$:$}}\rangle , 
\end{aligned} 
\end{gathered} 
\label{eq:97RQ}\end{gather}%
etc. A detailed parallelism between classical and quantum response characterisations of a solitary device is evident. It is worthy of a remark that, while \mbox{Eqs.\ (\ref{eq:30AN})} and (\ref{eq:47TT}) are equivalent, physical transparency of the former compares favourably to obscure formality of the latter.

\section{Electromagnetic self-action}%
\label{ch:B}%
Consider now the self-action problem for a solitary device. In quantum electrodynamics, it is expressed by the {\em dressing formula\/} \cite{Maxwell}, relating averages of the Heisenberg\ operator \mbox{$\protect{\hat{\mathcal J}}(t)$} to those of the ``free'' current \mbox{$\hat J(t)$}, 
\begin{align} 
\Phi_{\mathrm{m}} \big[
\zeta \big|A_{\mathrm{e}}
 \big] %
= 
\exp \bigg(
-i\frac{\delta }{\delta A_{\mathrm{e}}}G_{\text{R}}
\frac{\delta }{\delta \zeta }
 \bigg)
\Phi_{\mathrm{m}}^{\mathrm{I}} \big[
\zeta \big| A_{\mathrm{e}}
 \big] . 
\label{eq:54ZF}\end{align}%
We use condensed notation (\ref{eq:76NU}). 
Substituting (\ref{eq:24ST}) into the dressing formula we find, 
\begin{align} 
 & \Phi _{\mathrm{m}}
\protect{ [ \zeta | A_{\mathrm{e}} ]} \nonumber\\&\quad = \exp
 \Big ( -i\frac{\delta }{\delta A_{\mathrm{e}}}G_{\text{R}} \frac{\delta }{\delta \zeta } \Big ) \int \mathcal{D}[\mathcal{J}]
\,p^{\mathrm{I}}\protect{ [ \mathcal{J}| A_{\mathrm{e}} ]}\exp
 ( i\zeta \mathcal{J}) \nonumber\\&\quad = \int \mathcal{D}[\mathcal{J}]
\exp
 \Big ( \frac{\delta }{\delta A_{\mathrm{e}}}G_{\text{R}} \mathcal{J}\Big ) p^{\mathrm{I}}\protect{ [ \mathcal{J}| A_{\mathrm{e}} ]}\exp
 ( i\zeta \mathcal{J}) \nonumber\\&\quad = \int \mathcal{D}[\mathcal{J}]\,
p^{\mathrm{I}}\protect{ [ \mathcal{J}| A_{\mathrm{e}}+G_{\text{R}} \mathcal{J}]}\exp
 ( i\zeta \mathcal{J}) . 
\label{eq:43BB}\end{align}%
The last transformation here is an application of a functional shift operator, 
\begin{align} 
\exp
 \Big ( \frac{\delta }{\delta A_{\mathrm{e}}}G_{\text{R}} \mathcal{J}\Big ) p^{\mathrm{I}}\protect{ [ \mathcal{J}| A_{\mathrm{e}} ]}= p^{\mathrm{I}}\protect{ [ \mathcal{J}| A_{\mathrm{e}}+G_{\text{R}} \mathcal{J}]}. 
\label{eq:44BC}\end{align}%
The dressing formula (\ref{eq:54ZF}) is thus equivalent to the {\em quasistochastic dressing formula\/}, which is the stochastic dressing formula (\ref{eq:38AW}) regarded as a relation for P-functionals. 

Two remarks are in order here. Firstly, the dressing formula expresses the nontrivial part of perturbative calculations. In full-space problems, this is associated with suppression of infinities, which in turn are due to instantaneous self-actions. Further discussion of this issue may be found in Appendix \ref{ch:Reg}. Secondly, the very possibility to introduce quasidistribution \mbox{$p^{\mathrm{I}}$} as a characterisation of a bare device is yet another manifestation of the concept of obscured macroscopic view in the formal structure of quantum electrodynamics. Microscopic dynamics is screened from the ``electromagnetic\ observer'' from the start, by mere assumption that interaction with the quantised electromagnetic\ field is facilitated by the current. This happens {\em before\/} the field and interaction are actually introduced.

\section{Electromagnetic interaction of distinguishable devices}%
\label{ch:91RJ}%
\subsection{Preliminaries}%
\label{ch:48TU}%
The critical step to physics is from a solitary device to an interacting pair (\mbox{Fig.\ \ref{fig:BirdsEye14}}b). 
\mbox{Equations (\ref{eq:97RQ})} allow one to confine the discussion to the quantum current. The ``dressed'' and ``bare'' P-functionals for devices $A$ and $B$ and the corresponding characteristic functionals are introduced extending definitions (\ref{eq:20SP}), (\ref{eq:41AZ}) and (\ref{eq:21SQ}), (\ref{eq:24ST}) to the models of devices (\ref{eq:47MY}) and (\ref{eq:48MZ}). In practice, this boils down to assigning labels $A$ and $B$ to everything, e.g.,
\begin{align} 
\Phi _{\mathrm{m}A}
\protect{ [ \zeta | A_{\mathrm{e}} ]} & = 
\protect \langle {\mathcal T}{\mbox{\rm\boldmath$:$}}\exp
 ( i\zeta \protect{\hat{\mathcal J}}) {\mbox{\rm\boldmath$:$}}\rangle _A 
 \nonumber\\& = \int \mathcal{D}[\mathcal{J}]\, p_A[\mathcal{J}|A_{\mathrm{e}}]\exp
 ( i\zeta \mathcal{J}) . 
\label{eq:27SW}\end{align}%
The symbol \mbox{$\protect \langle \cdots \rangle _A$} denotes averages defined in the model of solitary device $A$, cf.\ \mbox{Eq.\ (\ref{eq:47MY})}. Similar definitions apply to other ``labeled'' quantities. 
\subsection{From {quasiprobability distribution}s to generating functionals}%
\label{ch:MG}%
As the first step to verification of \mbox{Eqs.\ (\ref{eq:5KS})} and (\ref{eq:8KV}) as quantum relations for P-functionals we rewrite them in terms of the corresponding characteristic functionals. Substituting the former into (\ref{eq:24ST}) we get, 
\begin{align} 
\Phi^{\mathrm{I}}_{\mathrm{m}}\protect{ [ 
\zeta 
 | 
A_{\mathrm{loc}} 
 ]} & = \int \mathcal{D}[\mathcal{J}_A] \mathcal{D}[\mathcal{J}_B] 
\exp[i\zeta (\mathcal{J}_A+\mathcal{J}_B)] 
 \nonumber\\&\quad \times p^{\mathrm{I}}_A[\mathcal{J}_A | A_{\mathrm{loc}}]p^{\mathrm{I}}_B[\mathcal{J}_B | A_{\mathrm{loc}}]
 . 
\label{eq:9KW}\end{align}%
The integral factorises, and we arrive at, 
\begin{align} 
\begin{aligned} 
 & \Phi_{\mathrm{m}} ^{\mathrm{I}} [ 
\zeta | A_{\mathrm{loc}}
 ] = 
\Phi_{\mathrm{m}A} ^{\mathrm{I}} [ 
\zeta | A_{\mathrm{loc}}
 ] \Phi_{\mathrm{m}B} ^{\mathrm{I}} [ 
\zeta | A_{\mathrm{loc}}
 ] . 
\end{aligned} 
\label{eq:63BF}\end{align}%
\mbox{Equation (\ref{eq:8KV})} takes more work. Substituting it into (\ref{eq:41AZ}) yields, 
\begin{align} 
\begin{aligned} 
\Phi_{\mathrm{m}} [ 
\zeta 
\big | 
A_{\mathrm{e}} 
 ] & = \int \mathcal{D}[\mathcal{J}_A] \mathcal{D}[\mathcal{J}_B] 
\exp[i\zeta (\mathcal{J}_A+\mathcal{J}_B)] 
 \\&\quad \times 
p_{A} [ 
\mathcal{J}_A \big| A_{\mathrm{e}} + G_{\text{R}} \mathcal{J}_B
 ] p_{B} [ 
\mathcal{J}_B \big| A_{\mathrm{e}} + G_{\text{R}} \mathcal{J}_A
 ] . 
\end{aligned} 
\label{eq:10KX}\end{align}%
We then take the route as in \mbox{Eq.\ (\ref{eq:43BB})}, but in reverse. We start from writing,
\begin{align} 
 & p_{A} [ 
\mathcal{J}_A \big| A_{\mathrm{e}} + G_{\text{R}} \mathcal{J}_B
 ] %
= \exp \bigg ( 
\frac{\delta }{\delta A_{\mathrm{e}}} G_{\text{R}} \mathcal{J}_B
 \bigg ) p_{A} [ 
\mathcal{J}_A \big| A_{\mathrm{e}}
 ] , 
\label{eq:11KY}\end{align}%
then pull the exponential differential operator out of the integral. Similar manipulations are applied to \mbox{$p_B$}. As a result we find, 
\begin{align} 
\Phi_{\mathrm{m}} [ 
\zeta 
\big | 
A_{\mathrm{e}} 
 ] & = \exp \bigg ( 
-i\frac{\delta }{\delta A_{\mathrm{e}}'} G_{\text{R}} \frac{\delta }{\delta \zeta } 
-i\frac{\delta }{\delta A_{\mathrm{e}}} G_{\text{R}} \frac{\delta }{\delta \zeta' }
 \bigg ) \nonumber\\&\quad \times 
\int \mathcal{D}[\mathcal{J}_A] \mathcal{D}[\mathcal{J}_B] 
\exp ( 
i\zeta \mathcal{J}_A + i\zeta' \mathcal{J}_B
 ) \nonumber\\&\qquad \times 
p_{A} [ 
\mathcal{J}_A \big| A_{\mathrm{e}}
 ] p_{B} [ 
\mathcal{J}_B \big| A_{\mathrm{e}}'
 ] |_{\zeta '=\zeta ,A_{\mathrm{e}}'=A_{\mathrm{e}}} . 
\label{eq:14LB}\end{align}%
Introducing pairs of variables $\zeta(t) ,\zeta '(t)$ and $A_{\mathrm{e}}(t),A_{\mathrm{e}}'(t)$ allows all differentiations to hit the right targets. The integral in (\ref{eq:14LB}) is factorised. Recalling (\ref{eq:27SW}) we arrive at the relation sought, 
\begin{align} 
\Phi_{\mathrm{m}} [ 
\zeta 
\big | 
A_{\mathrm{e}} 
 ] & = \exp \bigg ( 
-i\frac{\delta }{\delta A_{\mathrm{e}}'} G_{\text{R}} \frac{\delta }{\delta \zeta } 
-i\frac{\delta }{\delta A_{\mathrm{e}}} G_{\text{R}} \frac{\delta }{\delta \zeta' }
 \bigg ) \nonumber\\&\quad \times 
\Phi_{\mathrm{m}A} [ 
\zeta 
\big | 
A_{\mathrm{e}} 
 ] \Phi_{\mathrm{m}B} [ 
\zeta' 
\big | 
A_{\mathrm{e}}'
 ] \settoheight{\auxlv}{$ |$}%
\raisebox{-0.3\auxlv}{$ |_{\zeta '=\zeta ,A_{\mathrm{e}}'=A_{\mathrm{e}}}$}. 
\label{eq:66BK}\end{align}%
\subsection{Direct derivation of \mbox{Eq.\ (\ref{eq:63BF})} and (\ref{eq:66BK})}%
\label{ch:MT}%
\mbox{Equation (\ref{eq:63BF})} follows from factorisation of quantum averages for two distinguishable noninteracting subsystems, cf.\ \mbox{Sec.\ \ref{ch:29SY}}. To prove it, use the definition of \mbox{$\Phi_{\mathrm{m}} ^{\mathrm{I}}$} by \mbox{Eq.\ (\ref{eq:62ZQ})}, and the factorisation assumption (\ref{eq:84VJ}). We regard \mbox{Eq.\ (\ref{eq:63BF})} verified. In turn, this proves \mbox{Eq.\ (\ref{eq:5KS})} as a formula for P-functionals.

Substituting (\ref{eq:63BF}) into the dressing formula (\ref{eq:54ZF}) we obtain, 
\begin{align} 
\Phi_{\mathrm{m}} [ 
\zeta \big| A_{\mathrm{e}}
 ] & = 
\exp \bigg ( 
-i\frac{\delta }{\delta A_{\mathrm{e}}} G_{\text{R}} \frac{\delta }{\delta \zeta }
 \bigg ) \nonumber\\&\quad \times 
\Phi_{\mathrm{m}A} ^{\mathrm{I}} [ 
\zeta \big| A_{\mathrm{e}}
 ] \Phi_{\mathrm{m}B} ^{\mathrm{I}} [ 
\zeta \big| A_{\mathrm{e}}
 ] . 
\label{eq:95FM}\end{align}%
We now apply the relation \cite{Corresp}, 
\begin{align} 
\mathcal{F}_1 \bigg ( 
\frac{\delta }{\delta f}
 \bigg ) \mathcal{F}_2 [ 
f
 ] \mathcal{F}_3 [ 
f
 ] = \mathcal{F}_1 \bigg ( 
\frac{\delta }{\delta f}+
\frac{\delta }{\delta f'}
 \bigg ) \mathcal{F}_2 [ 
f
 ] \mathcal{F}_3 [ 
f'
 ] |_{f'=f} , 
\label{eq:96FN}\end{align}%
where $\mathcal{F}_1[\cdot]$, $\mathcal{F}_2[\cdot]$, $\mathcal{F}_2[\cdot]$ are arbitrary functionals and $f(t),f'(t)$ are auxiliary functional variables. Equation (\ref{eq:96FN}) is a compact way of formulating general rules of product differentiation. It may be verified expanding $\mathcal{F}_1[\cdot]$, $\mathcal{F}_2[\cdot]$, $\mathcal{F}_2[\cdot]$ in functional Taylor series. Using (\ref{eq:96FN}) we rewrite (\ref{eq:95FM}) as, 
\begin{align} 
\Phi_{\mathrm{m}} [ 
\zeta \big| A_{\mathrm{e}}
 ] & = 
\exp \bigg [ -i 
 \bigg ( 
\frac{\delta }{\delta A_{\mathrm{e}}} + 
\frac{\delta }{\delta A_{\mathrm{e}}'} 
 \bigg ) G_{\text{R}} \bigg ( 
\frac{\delta }{\delta \zeta } + 
\frac{\delta }{\delta \zeta '}
 \bigg ) \bigg ] \nonumber\\&\quad \times 
\Phi_{\mathrm{m}A} ^{\mathrm{I}} [ 
\zeta \big| A_{\mathrm{e}}
 ] \Phi_{\mathrm{m}B} ^{\mathrm{I}} [ 
\zeta' \big| A_{\mathrm{e}}'
 ] \settoheight{\auxlv}{$\big|$}%
\raisebox{-0.3\auxlv}{$\big|_{\zeta '=\zeta , 
A_{\mathrm{e}}'= A_{\mathrm{e}}}$}. 
\label{eq:97FP}\end{align}%
Expanding the bilinear form in the exponent we have, 
\begin{align} 
 & \exp \bigg [ -i 
 \bigg ( 
\frac{\delta }{\delta A_{\mathrm{e}}} + 
\frac{\delta }{\delta A_{\mathrm{e}}'} 
 \bigg ) G_{\text{R}} \bigg ( 
\frac{\delta }{\delta \zeta } + 
\frac{\delta }{\delta \zeta '}
 \bigg ) \bigg ] \nonumber\\&\quad = 
\exp \bigg ( 
-i\frac{\delta }{\delta A_{\mathrm{e}}'} G_{\text{R}} \frac{\delta }{\delta \zeta }
 \bigg ) \exp \bigg ( 
-i\frac{\delta }{\delta A_{\mathrm{e}}} G_{\text{R}} \frac{\delta }{\delta \zeta '}
 \bigg ) \nonumber\\&\qquad \times 
\exp \bigg ( 
-i\frac{\delta }{\delta A_{\mathrm{e}}} G_{\text{R}} \frac{\delta }{\delta \zeta }
 \bigg ) \exp \bigg ( 
-i\frac{\delta }{\delta A_{\mathrm{e}}'} G_{\text{R}} \frac{\delta }{\delta \zeta '}
 \bigg ) . 
\label{eq:98FQ}\end{align}%
The last two factors here ``dress'' the devices, 
\begin{align} 
\begin{aligned} 
\exp \bigg ( 
-i\frac{\delta }{\delta A_{\mathrm{e}}} G_{\text{R}} \frac{\delta }{\delta \zeta }
 \bigg ) \Phi_{\mathrm{m}A}^{\mathrm{I}} [ 
\zeta \big | A_{\mathrm{e}}
 ] & = 
\Phi_{\mathrm{m}A} [ 
\zeta \big | A_{\mathrm{e}}
 ] , \\ 
\exp \bigg ( 
-i\frac{\delta }{\delta A_{\mathrm{e}}'} G_{\text{R}} \frac{\delta }{\delta \zeta' }
 \bigg ) \Phi_{\mathrm{m}B}^{\mathrm{I}} [ 
\zeta' \big | A_{\mathrm{e}}'
 ] & = 
\Phi_{\mathrm{m}B} [ 
\zeta' \big | A_{\mathrm{e}}'
 ] , 
\end{aligned} 
\label{eq:65BJ}\end{align}%
cf.\ \mbox{Eq.\ (\ref{eq:54ZF})}, and we arrive at \mbox{Eq.\ (\ref{eq:66BK})}. In turn, this proves \mbox{Eq.\ (\ref{eq:8KV})} as a formula for P-functionals. 
\subsection{Extension to many devices}%
\label{ch:50TW}%
Extension to many devices is achieved by applying \mbox{Eqs.\ (\ref{eq:5KS})} and (\ref{eq:8KV}) recursively. So, for a set-up of $N$ devices, labeled by their number in place of $A$ and $B$, we obtain, 
\begin{align} 
p^{\mathrm{I}} [ 
\mathcal{J}| A_{\mathrm{loc}}
 ] & = \int \prod_{k=1}^N 
 \big \{ \mathcal{D}[\mathcal{J}_k] \,p^{\mathrm{I}}_k[\mathcal{J}_k | A_{\mathrm{loc}}] \big \} \nonumber\\&\quad \times 
\delta[\mathcal{J}_1+\cdots+\mathcal{J}_N -\mathcal{J}]
, 
\label{eq:51TX}\end{align}%
and 
\begin{align} 
p [ 
\mathcal{J}| A_{\mathrm{e}}
 ] & = \int \prod_{k=1}^N 
 \big \{ \mathcal{D}[\mathcal{J}_k] \,p_k[\mathcal{J}_k | A_{\mathrm{e}}+G_{\text{R}} (\mathcal{J}-\mathcal{J}_k)] \big \} \nonumber\\&\quad \times 
\delta [\mathcal{J}_1+\cdots+\mathcal{J}_N -\mathcal{J}]
, 
\label{eq:52TY}\end{align}%
For the corresponding characteristic functionals we have, 
\begin{align} 
 & \Phi_{\mathrm{m}} ^{\mathrm{I}} [ 
\zeta | A_{\mathrm{loc}}
 ] = \prod_{k=1}^N
\Phi_{\mathrm{m}k} ^{\mathrm{I}} [ 
\zeta | A_{\mathrm{loc}}
 ] , 
\label{eq:54UA}\end{align}%
and 
\begin{align} 
\Phi_{\mathrm{m}} [ 
\zeta 
\big | 
A_{\mathrm{e}} 
 ] & = 
 \bigg \{ \exp \bigg ( 
-i\sum_{l\neq n}\frac{\delta }{\delta A_{\mathrm{e}l}} G_{\text{R}} \frac{\delta }{\delta \zeta_n }
 \bigg ) \nonumber\\&\quad \times 
\prod_{k=1}^N
\Phi_{\mathrm{m}k} [ 
\zeta_k | A_{\mathrm{e}k}
 ] \bigg \} \Big|_{
 \{ \zeta,A_{\mathrm{e}} \} } , 
\label{eq:55UB}\end{align}%
where \mbox{$ \{ \zeta,A_{\mathrm{e}} \} $} stands for the substitution, 
\begin{align} 
\zeta _1=\cdots =\zeta _N=\zeta, \quad A_{\mathrm{e}1}=\cdots = A_{\mathrm{e}N}=A_{\mathrm{e}} . 
\label{eq:48BH}\end{align}%
These formulae generalise \mbox{Eqs.\ (\ref{eq:63BF})} and (\ref{eq:66BK}). They may be verified either applying them recursively, or adapting the method of \mbox{Sec.\ \ref{ch:MT}} to $N$ devices. 

Recursive nature of \mbox{Eqs.\ (\ref{eq:51TX})}--(\ref{eq:48BH}) reflects design of real macroscopic (e.g., electronic) devices. The latter are made of a large number of components, which, in our terms, are themselves electromagnetic\ devices 
(cf.\ the ``engineering viewpoint'' in \mbox{Sec.\ \ref{ch:85NC}}). True quantum mechanics only starts when this kind of subdivisibility stops.

\begin{center}* * *\end{center}

\mbox{Equations (\ref{eq:51TX})}--(\ref{eq:48BH}) give a formal expression to the statement made in the intro, that, {\em under macroscopic conditions, quantum electrodynamics\ and classical stochastic electrodynamics\ have identical dynamical structure.\/} One may also say that {\em flow of quantum information in a macroscopic setup is governed by laws of classical stochastic electrodynamics\/}. Truly quantum dynamics is limited to microscopic conditions of indistinguishability and/or equations of motion of matter.

\section{Response picture: the formal {\em versus\/} the physical}%
\label{ch:S}%
\subsection{Formal techniques revisited%
\label{ch:62UK}}%
As was explained in \mbox{Sec.\ \ref{ch:64UM}}, the way external c-number sources are employed in the response formulation is unphysical. This raises two questions: whether one can introduce response formulation without relying on sources, and whether one can reintroduce sources as a consistent macroscopic approximation. Both questions are answered in affirmative in this section. We also take this opportunity to reiterate the key technical points of our approach, with the emphasis on their interplay with the approximations. 

The actual calculations underlying our resuts utilize the functional aprroach to the Schwinger-Perel-Keldysh\ closed-time-loop\ techniques of quantum field theory\ \cite{SchwingerC,Perel,Keldysh} (for a recent review see \cite{KamenevLevchenko}). This is the minimal framework that allows one to include such concepts as {\em arbitrary macroscopic electromagnetic\ device\/} in a dynamical approach. 

The aforementioned key technical points of our approach are, 
\begin{itemize}
\item
{\em Response transformation\/} of the closed-time-loop\ techniques \cite{API,APII,APIII,WickCaus,DirResp,Maxwell}, prompted by the wave quantisation formula (\ref{eq:40MR}). 
\item
{\em Wick's theorem\/} in its functional (Hori's) form \cite{Wick,Hori,VasF}.
\end{itemize}
Both ``key technical points'' are exact and formally independent, but even a cursory inspection reveals that they are deeply connected. Indeed, consider the problem with the Hamiltonian (cf.\ \cite{API}), 
\begin{align} 
\hat H_{\text{I}} (t) = - \hat A(t)J_{\mathrm{e}}(t) . 
\label{eq:70UT}\end{align}%
It is easily solved, 
\begin{align} 
\protect{\hat{\mathcal A}}(t) = \hat A(t) + G_{\text{R}} J_{\mathrm{e}}(t) , 
\label{eq:71UU}\end{align}%
cf.\ \mbox{Eq.\ (\ref{eq:41MS})}.
The no-backaction nature of a c-number source automatically makes the electromagnetic interaction\ directed. The question whether directedness may also be introduced in nonlinear problems is answered in the affirmative by the wave quantisation formula (\ref{eq:40MR}). In turn, applying response transformation to Wick's theorem reveals its profound association with response properties of free quantised fields \cite{WickCaus}. 

\mbox{Equation (\ref{eq:70UT})} is the simplest example of strict causality in the response of a quantum system to sources \cite{API,APII,APIII,RelCaus,RelCausMadrid}. ``Strict causality'' means that it holds on all scales, macroscopic as well as microscopic. It is the``strict causality'' that makes the response formulation intuitive. Paradoxical as it may sound, it is also the ``strict causality'' that makes it unphysical, because it implies one's ability to vary sorces arbitrarily on microscopic scales. 
Application of the responce formulation to a solitary electromagnetic\ device \cite{DirResp,Maxwell} and its extention to electromagnetic interactions of distinguishable devices (\mbox{Sec.\ \ref{ch:91RJ}} of this paper) preserve both its advantages (causality and intuitiveness) and its main drawback (unphysical nature of c-number souces). 

\begin{figure}[b]
\begin{center}
\includegraphics[scale=0.4]{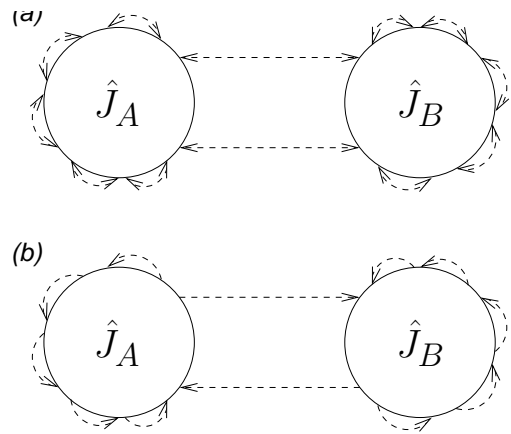}
\end{center}
\caption{The two-device model without external c-number sources. Dashed lines with arrows symbolise electromagnetic interaction. Circles surrounded by curved dashed arrows symbolise ``dressed'' devices regarded as macroscopic entities.
(a) Separation of interactions from self-actions emerges as a consequence of distinguishability. At this stage, electromagnetic\ actions and back-actions are not formally distinguished. This is emphasized by making all ``electromagnetic\ arrows'' bidirectional.
(b) The same with propagating waves introduced through the wave quantisation formula (\ref{eq:40MR}). The ``electromagnetic\ arrows'' are directional as in \mbox{Fig.\ \ref{fig:BirdsEye14}}.}
\label{fig:BirdsEye10}\end{figure}
\subsection{Dressed macroscopic devices}%
\label{ch:BE}%
Can one rederive the results in a manner preserving the advantages but without the drawbacks? Such alternative derivation is the subject of this section. In the main body of the paper, we resort to a ``pictorial argument''. All calculations are relegated to the Appendices. 

The starting point of the alternative derivation is the two-device model (\ref{eq:46MX}) without sources,
\begin{align} 
A_{\mathrm{e}}(t) = 0. 
\label{eq:58UE}\end{align}%
One also has to change the order in which the ``key technical points'' (cf.\ \mbox{Sec.\ \ref{ch:62UK}}) are applied to the two-device problem. Firstly, construct a perturbative solution to this problem in the closed-time-loop\ framework with Wick's theorem in its conventional form, secondly, impose the two-device approximation, and only in the very end perform the response transformation. The first step of this schedule may be borrowed from preprint \cite{QDynResp}, by dropping sources in the closed perturbative formula (56). The other two steps are outlined in Appendix \ref{ch:2DC}. 

Dressed devices as scattering entities emerge in such model due to an interplay of distinguishability with the wave quantisation formula (\ref{eq:40MR}). Indeed, consider the formal structure of electromagnetic interaction\ {\em before\/} the response transformation (Fig.\ \ref{fig:BirdsEye10}a). This picture is a schematic representation of \mbox{Eq.\ (\ref{eq:73UW})} in the Appendix. Assumed distinguishability of the devices leads to separation of the electromagnetic self-action problems for the devices from the problem of their electromagnetic interaction. ``Separation'' means that the self-action problems for the devices may be formulated within models (\ref{eq:47MY}) and (\ref{eq:48MZ}) without any reference to the two-device model. 

Neither Fig.\ \ref{fig:BirdsEye10}a nor the underlying \mbox{Eq.\ (\ref{eq:73UW})} ``know'' anything about directedness of the electromagnetic interaction. Graphically, this is emphasized by making dashed arrows symbolising the electromagnetic interaction\ bidirectional. 
The concept of ``dressed macroscopic device'' relies only {\em on the assumption of distinguishability and on the formal structure of quantum electrodynamics\/}. It is not limited to any kind of response characterisation of a quantum system.

The structure of electromagnetic interactions {\em after\/} the response transformation is illustrated in Fig.\ \ref{fig:BirdsEye10}b. This picture is a schematic representation of \mbox{Eq.\ (\ref{eq:74UX})}. In response picture, free electromagnetic field\ is characterised by the retarded propagator \mbox{$D_{\text{R}} $}. Consequently all ``electromagnetic\ arrows'' in Fig.\ \ref{fig:BirdsEye10}b are directional as in \mbox{Fig.\ \ref{fig:BirdsEye14}}.
The way Fig.\ \ref{fig:BirdsEye10}b is drawn emphasises that the wave quantisation formula gives rise to characterisation of the individual devices as {\em dressed response entities\/}. This happens {\em irrespective of c-number sources\/}---which, we remind, are absent in the approach illustrated in \mbox{Fig.\ \ref{fig:BirdsEye10}}. 

\begin{figure}[b]
\begin{center}
\includegraphics[scale=0.4]{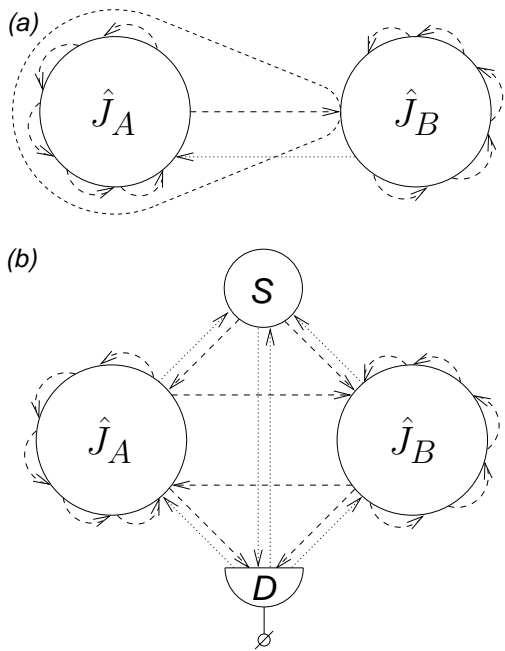}
\end{center}
\caption{C-number sources and macroscopic detectors as a macroscopic approximation. 
(a) The photodetection problem. The macroscopic back-action of the detector (dotted arrow) is neglected. The dashed oval isolates the source$+$field entity, operating independently of the detector. Formally, this entity is characterised by functional (\ref{eq:70QH}) in Appendix \ref{ch:PhDet}.
(b) The four-device arrangement, to which response characterisation of the two-device model (cf.\ \mbox{Fig.\ \ref{fig:BirdsEye14}}b) is an approximation. Neglected macroscopic actions and back-actions are shown as dotted arrows.%
}
\label{fig:BirdsEye11}\end{figure}

\subsection{Formal versus physical sources in macroscopic quantum electrodynamics\label{ch:61NP}}
The central formal result of this paper is that the dressed response entities in \mbox{Fig.\ \ref{fig:BirdsEye10}}b are identical with those emerging in formal models of the solitary devices with sources (\mbox{Fig.\ \ref{fig:BirdsEye14}}a). As formal concepts, these models were discussed in \mbox{Sec.\ \ref{ch:NP}}. The formal expression of this identity is the consistency condition.

Reintroducing sources in this approach starts from recognising the drastically different roles of single-directedness in the self-action and in the interaction problems. In the case of self-action, both ends of $ G_{\text{R}} $ are attached, so to speak, to the same electron. There is no way to control microscopic actions and back-actions associated with emitting and reabsorbing the field by matter at a truly microscopic level. Whereas in case of interaction, the signals travelling from device $A$ to device $B$ may be {\em experimentally distinguished\/} from those travelling from device $B$ to device $A$.

Consider, e.g., the generalised photodetection problem (\mbox{Fig.\ \ref{fig:BirdsEye11}}a). Calculations underlying this figure are relegated to Appendix \ref{ch:PhDet} (cf.\ also endnote \cite{endOSimpDet}). One starts from two scattering entities depicted in \mbox{Fig.\ \ref{fig:BirdsEye10}}b. The physical assumption turning them into {\em source\/} and {\em detector\/} is that the macroscopic electromagnetic\ backaction (shown in \mbox{Fig.\ \ref{fig:BirdsEye11}}a as a dotted arrow) can be neglected. This does not interfere with self-action problems for the source and detector, nor with the accuracy with which their properties may be accounted for. 

Neglecting macroscopic backaction gives rise to the concept of detected field. The corresponding formal entity is enclosed in \mbox{Fig.\ \ref{fig:BirdsEye11}}a in the dashed oval. Formally, the detected field is the radiated field in the model of the solitary device $A$. It is described by the time-normal\ field averages, 
\begin{align} 
\protect \langle {\mathcal T}{\mbox{\rm\boldmath$:$}}\protect{\hat{\mathcal A}}\cdots{\mbox{\rm\boldmath$:$}}\rangle _{A,0} . 
\label{eq:13SF}\end{align}%
The additional subscript 0 reminds that all c-number sources should be put to zero. 

In general, averages (\ref{eq:13SF}) cannot be mimicked by any classical phenomenology (quantum states of detected radiation). The semiclassical photodetection theory is recoved by assuming that such phenomenology (the {\em classical doppelganger\/} of the source, see \mbox{Sec.\ \ref{ch:56UC}} below) does exist. In this phenomenology one introduces a random classical field \mbox{$\mathcal{A}(t)$}, such that,
\begin{gather} 
\begin{gathered} 
\protect \langle \protect{\hat{\mathcal A}}(t) \rangle _{A,0} \approx 
\protect \langle \mathcal{A}(t) \rangle , \quad
\protect \langle {\mathcal T}{\mbox{\rm\boldmath$:$}}\protect{\hat{\mathcal A}}(t)\protect{\hat{\mathcal A}}(t'){\mbox{\rm\boldmath$:$}}\rangle _{A,0} \approx \protect \langle \mathcal{A}(t)\mathcal{A}(t') \rangle , 
\end{gathered} 
\label{eq:14SH}\end{gather}%
etc. This defines sources of radiation in a classical state. For coherent sources, averages (\ref{eq:13SF}) factorise, 
\begin{align} 
\protect \langle \protect{\hat{\mathcal A}}(t) \rangle _{A,0} \approx A_{\mathrm{e}}(t) , \quad 
\protect \langle {\mathcal T}{\mbox{\rm\boldmath$:$}}\protect{\hat{\mathcal A}}(t)\protect{\hat{\mathcal A}}(t'){\mbox{\rm\boldmath$:$}}\rangle _{A,0} \approx A_{\mathrm{e}}(t)A_{\mathrm{e}}(t') , 
\label{eq:15SJ}\end{align}%
etc. 

As far as the detection process is concerned, a coherent source is fully accounted for by the interaction with the external field \mbox{$A_{\mathrm{e}}$} in the Hamiltonian. Similar arguments apply to other arrangements with external sources. In particular, response characterisation of a two-device model (\mbox{Fig.\ \ref{fig:BirdsEye14}}b) is physically an approximation to the four-device arrangement (\mbox{Fig.\ \ref{fig:BirdsEye11}}b), where some of the macroscopic actions and backactions (shown as dotted arrows) have been neglected.

External sources introduced in this manner are ``coarse-grained'' quantities limited to macroscopic scales and conditions of distinguishability. They suffice for a macroscopic characterisation of response properties of devices. Upgrading this characterisation to a response formulation requires external sources placed inside atoms and electrons. This {\em physically unjustified\/} leap of logic recovers response formulation as a formal viewpoint, while also making it unphysical. In particular, theoretical response characterisation of a device is more detailed than anything one can measure in a real experiment. Recovering the former from the latter is always a matter of inference. 

Two remarks are in place here. Firstly, we called the photodetection arrangement in \mbox{Fig.\ \ref{fig:BirdsEye11}}a {\em generalised\/} because we did not require classicality of the output current of the detector. Such classicality is an additional assumption, distinguishing real photodetectors in the true meaning of the term (cf.\ \mbox{Sec.\ \ref{ch:60NN}} below). On the other hand, that this classicality is not needed allows one to easily generalise \mbox{Fig.\ \ref{fig:BirdsEye11}}a to {\em cascaded systems\/} \cite{CascadeC,CascadeH}. Secondly, in practice, control of macroscopic electromagnetic\ actions and back-actions is very much a matter of engineering. Examples are control of stray currents and fields in electronics and prevention of reflection of light toward laser sources in optics (to name just two). 
\section{Discussion: classically behaving quantum systems, classical world and quantum measurement}%
\label{ch:Dis}%
\subsection{Operational approach to quantum measurement}%
\label{ch:39QL}%
We take the liberty of starting this section with a somewhat lengthy quotation from W.\ Lamb's
renowned paper on operational approach to quantum mechanics \cite{LambOper}: 
\\
``A discussion of the interpretation of quantum mechanics on any level beyond [regarding it a set of rules for calculating physical properties of matter] almost inevitably becomes rather vague. The major difficulty involves the concept of measurement \mbox{$
\protect \langle 
\ldots
 \rangle $}. 
I have taught graduate courses in quantum mechanics for over 20 years \mbox{$
\protect \langle 
\ldots
 \rangle $}, and for almost all of them have dealt with measurement in the following manner. On beginning of lectures I told the students, `You must first learn the rules of calculation in quantum mechanics, and then I will tell you about the theory of measurement and discuss the meaning of the subject.'%
\mbox{$
\protect \langle 
\ldots
 \rangle $}
My attitude towards such problems has no doubt been influenced by contact with some research in experimental physics in which atomic states are ``manipulated'' by microwave and radiofrequency fields. 
In discussion of measurement of some dynamical variable of a physical system I want to know what apparatus is necessary for the task and how to use it, at least in principle.%
''

This paper follows Lamb's remarks, albeit more in spirit than to the letter. 
To begin with, it does not seem to be possible to introduce quantum measurement as a primary concept. The reason is the indistinguishability of microscopic objects (elementary particles). It is difficult to imagine how measurement could be discussed if one cannot tell the measured system from the measurement apparatus and from the observer. {\em Measurement is limited to macroscopic interactions under conditions of distinguishability.\/} Quantum measurement as a formal concept may therefore only emerge as an {\em approximation\/} within a fully developed theory of microscopic interactions---``rules of calculation'' first, ``measurement'' second.

Furthermore, quantum measurement is much more dependent on fundamental interactions available in nature, and on limitations of the human observer, than it is commonly acknowleged. The only kind of interaction available to human beings without use of auxiliary quantum systems as measurement apparata is the electromagnetic\ one. It is a long-range interaction, which acts at distances where all short-range interactions may be neglected. The latter remain hidden inside macroscopic devices, being responsible for their observable properties. It is the chierarchy of characteristic interaction lengths in nature that is ultimately responsible for the very existence of the macroscopic world, as perceived by a human observer.

\subsection{Mesoscopic coordinates of quantum systems and the ``obscured macroscopic view''}%
\label{ch:40QM}%
The critical distinction to be made in a quantum measurement theory is between {\em the observed\/} and {\em the inferred\/}. No particular macroscopic experiment gives full access to the underlying quantum dynamics (nor is this access necessary---think of quarks and gluons in quantum optics). One should therefore distinguish the directly {\em measured\/} mesoscopic quantities, and the microscopic ones, the properties of which may only be {\em inferred\/} from the experimental data (we remind that we treat ``mesoscopic'' and ``macroscopic'' as synonyms, cf.\ endnote \cite{endMeso}). 

We call measurable quantities {\em mesoscopic coordinates\/} of the quantum system in question. 
Real macroscopic experiments give access to properties of mesoscopic coordinates, thus allowing one only a fairly obscured and limited view of microscopic quantum dynamics. It is therefore not surprising that certain properties of mesoscopic coordinates (notably, those associated with propagation of quantum information in macroscopic experiments) turn out to be classical no matter what. 

Given the actual set of fundamental interactions in nature, mesoscopic coordinates of real devices may only be electromagnetic. However, the idea of separation of the macroscopic and microscopic levels of insight is not confined to quantum mechanics, nor to electrodynamics, nor is it anything new. It underlies thermodynamics, whether classical or quantum. It may also be found in any case where some kind of ``coarse-grained'' description is introduced (e.g., in hydrodynamics).

Sometimes our view of a particular quantum system is obscured to the extent that its quantum properties become fully hidden, and what we see looks perfectly classical. 
This was exactly the situation before first experimental 
indications of the underlying quantum world---properties of the 
black-body radiation and of the photoelectric effect---emerged 
in the experiment. 
We then say that the {\em seemingly\/} classical entity we are observing is the {\em classical doppelganger\/} of the quantum system in question. By definition, such classicality is limited to properties of mesoscopic coordinates: an electron cannot look classical while an electron current can. A {\em classically behaving quantum system\/} is a system that has a classical doppelganger; in other words, it is a system that {\em appears\/} classical in some experiment. A system that does not behave classically is said to exhibit quantum behaviour ({\em quantum-behaved systems\/}).

An immediate word of caution is in place here. As mentioned in the introduction, we associate ``macroscopic'' and ``microscopic'' with ``distinguishable'' and ``indistinguishable'', and not with ``large'' and ``small''. For example, electrons in an EPR pair remain indistinguishable until environmental influences, such as collisions with backgroud gas, destroy their entanglement. The EPR pair must be regarded a microscopic object, irrespective of how large it becomes over its lifetime (there is more to this than mere terminology, see below). On the other hand, in a hydrogen atom, the electron and proton are distinguishable unless pair creation needs to be taken into account. This allows one to go as far in the theory as the leading (Bethe) contribution to the Lamb shift \cite{Schweber} (cf.\ also endnote \cite{endHResp}). This example also shows that observable quantum effects have two origins: properties of individual quantum particles and their indistinguishability.

In this paper, we do not include collisions of distinguishable matter subsystems. ``Collision'' is any situation where conditions of distinguishability in a setup vary with time. Collisions of elementary particles are obviously covered by this definition. Whether one regards the EPR pair ``microscopic'' or ``macroscopic'' depends on whether collisions are explicitly accounted for. For purposes of this paper, collisions belong to microscopic dynamics. Integration of the latter into our approach remains subject to further work (there does not seem to be any formal obstacle to that, cf.\ \cite{BWO,APIII,DirResp,Maxwell}).

One more reservation is in order here. That only the electromagnetic\ current is observed in real experiments is a limitation of humans, and not a property of quantum mechanics. A creature with the sixth sense of strong interaction would see the world very differently, and require a very different measurement theory. 

\subsection{The classically behaving devices and the classical world}%
\label{ch:56UC}%
In \mbox{Sec.\ \ref{ch:40QM}}, we defined a classically behaving quantum system\ as a system that has a classical doppelganger. In quantum electrodynamics, this implies nonnegativity of the relevant quasiprobability distribution, allowing one to formally interpret it as a probability distribution. So, device $A$ in \mbox{Fig.\ \ref{fig:BirdsEye14}} is classically behaving, if the P-functional characterising it is nonnegative, 
\begin{align} 
p_A[\mathcal{J}|A_{\mathrm{e}}] \geq 0 . 
\label{eq:57UD}\end{align}%
This does not change the fact that this device is a quantum system, nor does it in any way affect validity of the general quantum formula for P-functionals (\ref{eq:52TY}). Classically behaving devices interact with the world according to laws of quantum, not classical, electrodynamics. That under macroscopic conditions these two in essence coincide is a {\em property\/} of the conventional quantum electrodynamics.

Continuing this logic, ``classical world'' emerges when all devices within reach of the observer are classically behaving. The whole world looked classical to the human observer before first experimental 
indications of the underlying quantum world---properties of the 
black-body radiation and of the photoelectric effect---emerged 
in the experiment. 

The above definition of a classically behaving device\ calls for an extensive comment. 
This definition obviously depends on what we term ``quantum''. Formal techniques naturally puts the boundary at what we can/cannot {\em describe\/} in terms of classical stochastic electrodynamics; this is all what \mbox{Eq.\ (\ref{eq:57UD})} is about. According to such definition, a thermal source of radiation behaves classically (unlike, e.g., a KDP crystal). This is in agreement with the general distinction between the observed and the inferred. Indeed, quantum nature of the thermal radiation is recognised only so far as one attempts to {\em explain\/} it in some sort of classical dynamical model, and realises futility of all such attempts due to the equidistribution theorem. 

An important reservation is that the same real system may behave classically in some experiments and exhibit quantum behaviour in others. So, in quantum optics, the output current of a photodetector is commonly treated as a classical stochastic process. A deeper insight reveals that macroscopic currents are antibunched \cite{YamamotoElAB}.

The classical doppelganger\ of a classically behaving device\ is a classical system introduced regarding \mbox{$p_A[\mathcal{J}|A_{\mathrm{e}}]$} (say) a conditional probability distribution. This definition is strictly phenomenological; a dynamical classical model of a classically behaving device\ is not bound to exist. An example is again the black-body radiation. In certain cases such model may be constructed; this mostly depends on the degree of insight. So, an ideal mirror may be introduced by postulating boundary conditions for Maxwell equations, but an attempt to fully explain properties of real mirrors in classical terms is doomed. 

As a rule, taking the limit \mbox{$\hbar \to 0$} of a classically behaving device\ does not recover its classical doppelganger. Indeed, such limit turns a quantum Hamilton evolution into a classical Hamilton one. The latter is not stochastic. Thus all dynamical noises (such as shot noise) will be lost. The noises associated with quantum averaging (e.g., thermal noise) may survive such limit but get distortet (cf.\ the thermal P-function of the harmonic oscillator {\em versus\/} thermal distribution of the classical oscillator \cite{Schleich}). Worse still, such limit may not exist, as, for instance, for the black-body radiation. While interpreting Planck's curve as a probability distribution makes perfect phenomenological sense. 

Actually, it is not even guarantied that a particular classically behaving device\ is subject to a semiclassical model. This again depends on the degree of insight. As a nontrivial example we point to the linear response theory of the Dirac vacuum in \cite{DirResp}. In the first nonvanishing approximation, vacuum polarisation may be calculated, with the necessary regularisation and renormalisation, solely as a property of free fermions. Corrections of higher order would involve quantised electromagnetic field\ as part of the dynamical model. 

\begin{figure}[b]
\begin{center}
\includegraphics[scale=0.4]{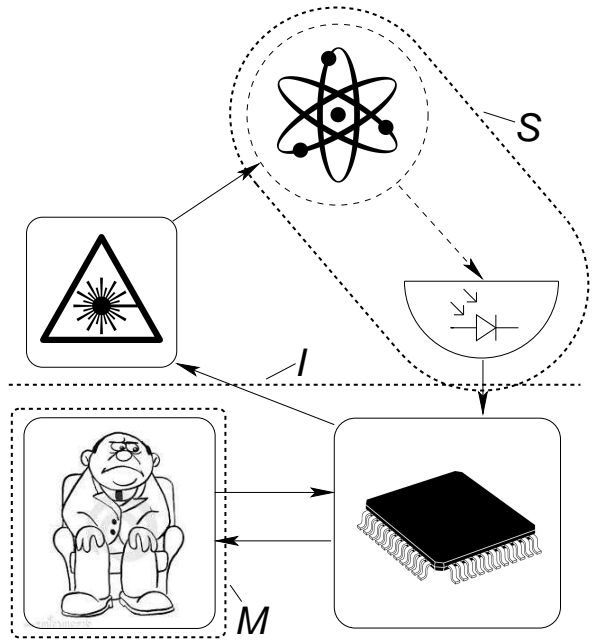}
\end{center}
\caption{Schematics of information flow in an optical experiment. Solid lines indicate ``the classical'', and dashed ones---``the quantum''. Thick dotted lines show possible choices of the quantum-classical boundary: ``minimal'' ($M$), ``intermediate'' ($I$), and ``strict'' ($S$).%
}
\label{fig:MacrExp}\end{figure}
\subsection{Electromagnetic quantum measurement}%
\label{ch:60NN}%
Equations (\ref{eq:51TX})--(\ref{eq:55UB}) allow one to mix classically and nonclassically behaving devices at will. However, not every such mix is a ``measurement''. The latter implies existence of a ``classical observer'', who is only capable of interacting with ``classical measurement apparata'' (cf.\ \mbox{Fig.\ \ref{fig:MacrExp}}). In physical terms, these are classically behaving devices, but this is not enough. For a set-up to become a measurement, it should exhibit the {\em quantum-classical boundary\/} and contain {\em macroscopic detectors.\/} Indeed, consider the flow of information in \mbox{Fig.\ \ref{fig:MacrExp}}. The observer interacts with a computer, which controls the laser source, and processes infomation coming from the photodetector. The latter detects radiation of the quantum system pumped by the laser. Devices encircled by solid lines are classically behaving; similarly, electromagnetic\ radiation in a classical state is depicted schematically by solid lines with arrows. The quantum system is encircled by dashed line; its radiation, which may be in a quantum state, is shown as dashed arrow.

Possible positions of the quantum-classical boundary\ in this set-up are shown by thick dotted lines. The ``minimal'' boundary separates the observer from the rest of the world, the ``intermediate'' one separates the observer and computer from the laser, detector and quantum system. The ``strict'' quantum-classical boundary\ separates the system and the detector from the rest the world---which in this case includes the observer. What is shared by all three is that the observer is always on one side of the boundary, while the system {\em and the detector\/}---on the other. 

For a classically behaving device\ to be a detector, its output radiation must be in a classical state irrespective of the state of the detected radiation. This is only possible if the detector adds noise of its own, such as shot noise of real photodetectors. The same requirement may be put as that the {\em compound device\/} comprising the quantum system and the detector be a classically behaving\ one. Treating the quantum system and the detector as an entity turns the measurement setup into a ``classical world''.

Positioning of possible quantum-classical\ boundaries in \mbox{Fig.\ \ref{fig:MacrExp}} depends on assumptions about the devices. If the laser radiation were in a quantum state, the strict quantum-classical boundary\ would shift to the ``intermediate'' one. In the future where the controlling device may be a quantum computer, the only possible choice of the quantum-classical boundary\ may become the ``minimal'' one. However, even in this future the observer will interact with what he perceives as a classically behaving device. 

\section{Conclusion}%
In conclusion, quantum theory of the electromagnetic\ interaction under macroscopic conditions of distinguishability of devices and of controlled actions and back-actions between them is constructed. This theory is shown to be subject to ``doing quantum electrodynamics\ while thinking classically'', allowing one to substitute essentally classical considerations for quantum ones without any loss in generality. Implications of these results for the quantum measurement theory are discussed.

\appendix 

\section{Self-action, causality and regularisations}%
\label{ch:Reg}%
As we have already mentioned, Eq.\ (\ref{eq:38AW}) is consistent only due to causality properties of the P-functionals. Here is a simple example. Assume that all quantities in (\ref{eq:38AW}) do not depend on time. The external field shifts the Gaussian distribution of the current, 
{\begin{align}{{
 \begin{aligned} 
p^{\mathrm{I}} \big(
 J \big | A_{\mathrm{e}} 
 \big) 
= \frac{1}{\sqrt{2\pi}\, J_0 }
\exp \bigg[
-\frac{ \big(
 J - \chi A_{\mathrm{e}} \big)^2 
}{2 J_0^2}
\bigg] , 
\end{aligned}}}%
\label{eq:3ZZ}\end{align}}%
where $ J_0>0$ and $\chi$ are real constants. In place of (\ref{eq:37AV}) we postulate a scalar formula, 
{\begin{align}{{
 \begin{aligned} 
 A_{\mathrm{loc}} = A_{\mathrm{e}} + G_{\text{R}} J, 
\end{aligned}}}%
\label{eq:4AA}\end{align}}%
where $ G_{\text{R}}$ is one more real constant. For the ``dressed current'' we find, 
{\begin{align}{{
 \begin{aligned} 
p \big(
 J \big | A_{\mathrm{e}} 
 \big) 
= \frac{1}{\sqrt{2\pi}\, J_0 }
\exp \bigg[
-\frac{ \big(
 J - \chi G_{\text{R}} J - \chi A_{\mathrm{e}} \big)^2 
}{2 J_0^2}
\bigg] . 
\end{aligned}}}%
\label{eq:5AB}\end{align}}%
This function is not normalised, 
{\begin{align}{{
 \begin{aligned} 
\int d J \, p \big(
 J \big | A_{\mathrm{e}} 
 \big) = \frac{1}{1-\chi G_{\text{R}}} \neq 1, 
\end{aligned}}}%
\label{eq:6AC}\end{align}}%
and cannot be a probability distribution for anything. 

To see how causality breaks this vicious circle of same-time interactions consider another simple example. We generalise (\ref{eq:3ZZ}) to two currents, 
\begin{multline}\hspace{0.4\columnwidth}\hspace{-0.4\twocolumnwidth} 
p^{\mathrm{I}} \big(
 J , J' \big | A_{\mathrm{e}}, A_{\mathrm{e}}' 
 \big) 
\\ 
= \frac{1}{2\pi J_0^2 }
\exp \bigg[
-\frac{
 \big( J - \chi A_{\mathrm{e}} \big)^2
+ 
 \big( J' - \chi A_{\mathrm{e}}' \big)^2 
}{2 J_0^2}
\bigg] . 
\hspace{0.4\columnwidth}\hspace{-0.4\twocolumnwidth} 
\label{eq:7AD}\end{multline}%
The primed current preceds the unprimed one in time; therefore it may affect the latter but not {\em vice versa\/}. Same-time interactions are not allowed either. The simplest case of such interaction is, 
{\begin{align}{{
 \begin{aligned} 
& A_{\mathrm{loc}} = A_{\mathrm{e}} + G_{\text{R}} J', & 
& A_{\mathrm{loc}}' = A_{\mathrm{e}}' . 
\end{aligned}}}%
\label{eq:10AH}\end{align}}%
For the dressed currents we then have, 
\begin{multline}\hspace{0.4\columnwidth}\hspace{-0.4\twocolumnwidth} 
p \big(
 J , J' \big | A_{\mathrm{e}}, A_{\mathrm{e}}' 
 \big) 
= \frac{1}{2\pi J_0^2 }
\\ \times
\exp \bigg[
-\frac{
 \big( J - \chi G_{\text{R}} J' - \chi A_{\mathrm{e}} \big)^2
+ 
 \big( J' - \chi A_{\mathrm{e}}' \big)^2 
}{2 J_0^2}
\bigg] . 
\hspace{0.4\columnwidth}\hspace{-0.4\twocolumnwidth} 
\label{eq:8AE}\end{multline}%
Unlike (\ref{eq:5AB}), this function is normalised, 
{\begin{align}{{
 \begin{aligned} 
\int d J'
\int d J\,p \big(
 J , J' \big | A_{\mathrm{e}}, A_{\mathrm{e}}' 
 \big) = 1 . 
\end{aligned}}}%
\label{eq:9AF}\end{align}}%
It is therefore a genuine two-dimensional conditional probability distribution for a correlated pair of currents. 

The order of integrations in (\ref{eq:9AF}) is chosen so as to make the result obvious. Indeed, (\ref{eq:8AE}) has the structure, 
{\begin{align}{{
 \begin{aligned} 
p \big(
 J , J' \big | A_{\mathrm{e}}, A_{\mathrm{e}}' 
 \big) = p \big(
 J\big | A_{\mathrm{e}} , J' 
 \big)p' \big(
 J' \big | A_{\mathrm{e}}' 
 \big) , 
\end{aligned}}}%
\label{eq:12AK}\end{align}}%
where 
{\begin{align}{{
 \begin{aligned} 
p \big(
 J\big | A_{\mathrm{e}} , J' 
 \big) & = \frac{1}{\sqrt{2\pi}\, J_0 }
\exp \bigg[
-\frac{
 \big( J - \chi G_{\text{R}} J' - \chi A_{\mathrm{e}} \big)^2
}{2 J_0^2}
\bigg] , \\ 
p' \big(
 J' \big | A_{\mathrm{e}}' 
 \big) & = \frac{1}{\sqrt{2\pi}\, J_0 }
\exp \bigg[
-\frac{
 \big( J' - \chi A_{\mathrm{e}}' \big)^2 
}{2 J_0^2}
\bigg] . 
\end{aligned}}}%
\label{eq:13AL}\end{align}}%
The later current is conditional on the earlier one and the external field. The earlier current is conditional only on the external field. 
Similar structures should emerge for any time sequence of currents irrespective of any detals of the interaction. The only requirement is that each current depends only on those preceding it in time, and not on itself. 

In real problems with continuous time, critical for cancellation of same-time interactions are {\em regularisations\/}. So, in \mbox{Ref.\ \cite{BWO}}, {\em causal regularisation\/} was applied to the retarded Green function of the emerging equation for phase-space amplitudes. This made noise sources present in the said equation independent of the amplitudes at the same time, resulting in the Ito calculus being chosen. The effect of causal regularisation is thus twofold: to introduce an infinitesimal delay into the phase-space equation, which is in essence time discretisation, and to prevent same-time interactions. A conceptual connection between the above simple examples and the causal regularisation is obvious. An example of causal regularisation applied to a relativistic problem is the linear response theory of the Dirac vacuum in \cite{DirResp}, see also \cite{Maxwell}. Further discussion of this issue is outside the scope of this paper.

\section{Quantum electrodynamics in response representation}%
\label{ch:QR}%
\subsection{Interaction, factors and signs}%
\label{ch:GZ}%
Here we reiterate the relevant results of \cite{Maxwell}. For purposes of this appendix, we assume \mbox{$\hat H_{\text{I}} $} to be, 
\begin{align} 
\hat H_{\text{I}} (t) & = - \hat J(t)\hat A(t) - \hat J(t)A_{\mathrm{e}}(t) - \hat A(t)J_{\mathrm{e}}(t). 
\label{eq:25AH}\end{align}%
While interaction (\ref{eq:36MM}) suffices for all physical purposes, interaction (\ref{eq:25AH}) leads to a more symmetric formulation.

In paper \cite{Maxwell} a covariant formulation was used. Adapting its results to interaction (\ref{eq:25AH}) requires a certain amount of care. For simplicity, we assume \mbox{$c=1$} (anyway $c$ must disappear in a nonrelativistic formulation). Equally trivial replacements apply to the arguments of functions. The obvious part of adapting \cite{Maxwell} thus is, 
\begin{align} 
c\to 1, \quad x ,\mu \to t, \quad 
\int d^4 x \sum_{\mu } \to \int dt.
\label{eq:57BS}\end{align}%
To identify more subtle replacements we compare formulae for the S-matrix. 
Dropping the sources, the S-matrix associated with interaction (\ref{eq:25AH}) is, 
\begin{align} 
\protect{\hat{\mathcal S}}= T_+\exp
\frac{i\hat J\hat A}{\hbar } . 
\label{eq:55BQ}\end{align}%
The corresponding formula in \cite{Maxwell} reads, (with \mbox{$c=1$})
\begin{align} 
\protect{\hat{\mathcal S}}= T_+\exp
\frac{\hat J\hat A}{i\hbar} ,
\label{eq:60BV}\end{align}%
where condensed notation (\ref{eq:3VS}) is redefined according to the covariant standard, 
\begin{align} 
fg & = \int d^4 x f^{\mu }( x ) g_{\mu }( x ) \nonumber\\& = -\int d^4x 
 [ {\mbox{\rm\boldmath$f$}}( x ){\mbox{\rm\boldmath$g$}}( x )-f_{0}( x ) g_{0}( x ) ] . 
\label{eq:56BR}\end{align}%
Changed sign of the exponent leads to redefinition of the retarded propagator, 
\begin{align} 
D_{\text{R}} \to G_{\text{R}} = - D_{\text{R}} . 
\label{eq:58BT}\end{align}%
This implies correspondences (\ref{eq:57BS}). For more details on sign and factor conventions in nonrelativistic and relatistic approaches see Appendix C.1 in \cite{DirResp}.

The simplest way of adapting the formulae of \mbox{Ref.\ \protect\cite{Maxwell}} is thus to change the metric signature from \mbox{$
 \{ +,-,-,- \} $} to \mbox{$
 \{ -,+,+,+ \} $}. This leads to replacements, 
\begin{align} 
fg\to -fg, \quad f D_{\text{R}} g \to - f G_{\text{R}} g, 
\label{eq:59BU}\end{align}%
for all occurences of these kinds of condensed notation. Combined with (\ref{eq:57BS}), this allows one to adapt any formula in \cite{Maxwell} to interaction (\ref{eq:25AH}). 

\subsection{Closed-time-loop formalism and response transformation}%
\label{ch:UD}%
We work in the {\em closed-time-loop\/} formalism of the quantum field theory \cite{SchwingerC,Perel,Keldysh}. The system is described by the double-time-ordered averages of the Heisenberg\ field and current operators, (\mbox{$m,m',m'',m'''$} are arbitrary integers)
\begin{align} 
\begin{aligned} 
 & \protect \langle T_-
\protect{\hat{\mathcal J}}(t_1)\cdots\protect{\hat{\mathcal J}}(t_m) 
\protect{\hat{\mathcal A}}(t'_1)\cdots\protect{\hat{\mathcal A}}(t'_{m'}) 
 \\&\quad \times
T_+\protect{\hat{\mathcal J}}(t''_1)\cdots\protect{\hat{\mathcal J}}(t''_{m''}) 
\protect{\hat{\mathcal A}}(t'''_1)\cdots\protect{\hat{\mathcal A}}(t'''_{m'''}) \rangle , 
\end{aligned} 
\label{eq:10BU}\end{align}%
where \mbox{$T_{\pm}$} are the forward and reverse time orderings (often denoted $T$ and $\bar T$).
By definition, bosonic operators commute under all kinds of ordering. We note in passing that definition (\ref{eq:10BU}) may cause mathematical problems, see the concluding remark in appendix A1 in \cite{WickCaus}. 

Averages of quantities (\ref{eq:10BU}) are conveniently accessed through their characteristic functional, 
\begin{align} 
 & \Phi [
\eta ,
\zeta 
\big | 
j_{\mathrm{e}},
a_{\mathrm{e}}
| 
J_{\mathrm{e}}, 
A_{\mathrm{e}} 
] 
 \\&\quad = \langle T_- 
\exp (
-i{\eta}_- {\hat{\mathcal E}} 
-i{\zeta}_- {\hat{\mathcal D}}
 ) T_+ 
\exp (
i{\eta}_+ {\hat{\mathcal E}} 
+i{\zeta}_+ {\hat{\mathcal D}} 
 ) \rangle
|_{\mathrm{c.v.}} . 
\label{eq:51ZC}\end{align}%
where 
$\eta_{\pm}(t)$ and 
$\zeta_{\pm}(t)$ 
are auxiliary complex c-number functions, and c.v.\ (short for {\em causal variables\/}) refers to the set of {\em response substitutions\/} \cite{API,APII}, 
\begin{align}
\eta_{\pm}(t) 
& = \frac{ j_{\mathrm{e}}(t)}{\hbar }\pm \eta ^{(\mp)}(t) 
, 
\label{eq:21FB}\\ 
\zeta_{\pm}
(t) 
 & 
= \frac{ a_{\mathrm{e}}
(t) 
}{\hbar }\pm \zeta ^{(\mp)}
(t) 
. 
\label{eq:23SS}\end{align}%
The symbols ${}^{(\pm)}$ denote separation of the frequency-positive and negative\ parts as per Eq.\ (\ref{eq:07PA}). The averaging in (\ref{eq:51ZC}) is over the initial (Heisenberg) state of the system (\ref{eq:40BJ}).

We remind that the initial state of the field is vacuum, cf. \mbox{Eq.\ (\ref{eq:42BL})}. In terms of \cite{Maxwell}, this means that \mbox{$A_{\mathrm{in}} =0$}. The corresponding argument of \mbox{$\Phi $} is redundant and has been omitted. Note also that the choice of signs in (\ref{eq:51ZC}) matches (\ref{eq:55BQ}). 
\subsection{Consistency conditions}%
\label{ch:UPC}%
The critical property of functional (\ref{eq:51ZC}) is that it depends only on sums of the external sources and the corresponding auxiliary variables, \mbox{$
J_{\mathrm{e}}+j_{\mathrm{e}}
$} and \mbox{$
A_{\mathrm{e}}+a_{\mathrm{e}}
$} \cite{Maxwell}, 
\begin{align} 
\Phi \big[
\eta ,\zeta
\big | 
j_{\mathrm{e}},
a_{\mathrm{e}}
\big | 
J_{\mathrm{e}},
A_{\mathrm{e}}
 \big] %
= 
\Phi \big[
\eta ,\zeta
\big | 
0,0 
\big | 
J_{\mathrm{e}}+j_{\mathrm{e}},
A_{\mathrm{e}}+a_{\mathrm{e}}
 \big] 
. 
\label{eq:29FL}\end{align}%
Alternatively, 
\begin{align} 
\Phi \big[
\eta ,\zeta
\big | 
j_{\mathrm{e}},
a_{\mathrm{e}}
\big | 
J_{\mathrm{e}},
A_{\mathrm{e}}
 \big] %
= 
\Phi \big[
\eta ,\zeta
\big | 
J_{\mathrm{e}}+j_{\mathrm{e}},
A_{\mathrm{e}}+a_{\mathrm{e}}
\big | 
0,0 
 \big] 
. 
\label{eq:38FV}\end{align}%
These relations are a generalisation of {\em consistency conditions\/} derived in \mbox{Refs.\ \cite{APII,APIII}}. 
\subsection{Response viewpoint and the time-normal ordering}%
\label{ch:RTN}%
Equations (\ref{eq:29FL}) and (\ref{eq:38FV}) express the same functional but much differ in their interpretation. Equation (\ref{eq:29FL}) introduces the formal {\em response viewpoint\/}, the one we mostly adhere to in this paper. Equation (\ref{eq:38FV}) shows that, mathematically, external c-number sources in quantum equations of motion are redundant: information contained in the {Heisenberg}\ operators conditional on the sources is already present in the operators defined without the sources. This allows one to reintroduce response viewpoint without relying on sources. For a discussion see \mbox{Sec.\ \ref{ch:S}}.

In view of Eq.\ (\ref{eq:29FL}) we may set the redundant auxiliary variables to zero, 
{\begin{align}{{
 \begin{aligned} 
a_{\mathrm{e}}(t) = 0, \quad j_{\mathrm{e}}(t) = 0. 
\end{aligned}}}%
\label{eq:24FE}\end{align}}%
This does not lead to any loss of generality (assuming the souces may be arbitrary, cf.\ \mbox{Sec.\ \ref{ch:S}}). Complete quantum formulae may be recovered replacing, 
{\begin{align}{{
 \begin{aligned} 
& A_{\mathrm{e}}\to a_{\mathrm{e}}+A_{\mathrm{e}}, & 
& J_{\mathrm{e}}\to j_{\mathrm{e}}+J_{\mathrm{e}} . 
\end{aligned}}}%
\label{eq:31FN}\end{align}}%
Under (\ref{eq:24FE}), the formal description of the system is given by the reduced characteristic functional, 
{\begin{align}{{
 \begin{aligned} 
\Phi \big[
\eta ,\zeta
\big | 
J_{\mathrm{e}},
A_{\mathrm{e}} 
 \big] 
= 
\Phi \big[
\eta ,\zeta
\big | 
0,0 
\big | 
J_{\mathrm{e}},
A_{\mathrm{e}}
 \big] 
. 
\end{aligned}}}%
\label{eq:77JN}\end{align}}%
We {\em postulate\/} this functional to be the characteristic one for the the time-normal averagea of the quantised current and field, 
\begin{align} 
\Phi \big[
\eta ,\zeta
\big | 
J_{\mathrm{e}},
A_{\mathrm{e}} 
 \big] = 
\protect \langle {\mathcal T}{\mbox{\rm\boldmath$:$}}\exp
 ( i\eta \protect{\hat{\mathcal A}}+ i \zeta \protect{\hat{\mathcal J}}) {\mbox{\rm\boldmath$:$}}\rangle . 
\label{eq:26AJ}\end{align}%
The Heisenberg\ operators depend on the sources by construction. 
Postulating (\ref{eq:26AJ}) recovers the beyond-the-RWA definition of the time-normal ordering\ \cite{APII}; cf.\ \mbox{Sec.\ \ref{ch:33TC}}.

\subsection{Reduction to current}%
\label{ch:UPD}%
Full electromagnetic properties of the quantum device may be expressed by the properties of the {Heisenberg}\ (``dressed'') current operator. The latter are contained in the functional, 
\begin{align} 
\Phi_{\mathrm{m}} \big[
\zeta 
\big | 
A_{\mathrm{e}} 
 \big] 
= \Phi \big[
0 ,\zeta
\big | 
0,
A_{\mathrm{e}} 
 \big] . 
\label{eq:52ZD}\end{align}%
Namely \cite{Maxwell}, 
\begin{align} 
\Phi \big[
\eta ,\zeta 
\big | 
J_{\mathrm{e}},
A_{\mathrm{e}} 
 \big] %
= \exp \big(
i\eta G_{\text{R}} J_{\mathrm{e}} 
 \big) 
\Phi_{\mathrm{m}} \big[
\zeta+\eta G_{\text{R}} 
\big | 
A_{\mathrm{ext}}
 \big] , 
\label{eq:53ZE}\end{align}%
where \mbox{$
A_{\mathrm{ext}}
$} is the natural combinations of the sources, 
{\begin{align}{{
 \begin{aligned} 
 A_{\mathrm{ext}}(t) & = A_{\mathrm{e}}(t) 
+ \int dt' G_{\text{R}}(t-t') J_{\mathrm{e}}(t')
. 
\end{aligned}}}%
\label{eq:26FH}\end{align}}%
With \mbox{$J_{\mathrm{e}}=0$}, \mbox{Eq.\ (\ref{eq:53ZE})} is nothing but a compact form of \mbox{Eqs.\ (\ref{eq:97RQ})}. 

\subsection{Dressing the current}%
\label{ch:25SU}%
Nontrivial part of perturbative calculations is expressed by the ``dressing formula'' (\ref{eq:54ZF}). The no-sources definitions of the functionals \mbox{$\Phi_{\mathrm{m}}$} and \mbox{$\Phi_{\mathrm{m}}^{\mathrm{I}}$} read, 
\begin{align} 
\Phi_{\mathrm{m}} [
\zeta | a_{\mathrm{e}}
 ] 
 & = 
\protect \langle 
T_-\exp (
-i{\zeta}_- \protect{\hat{\mathcal J}}) T_+\exp (
i{\zeta}_+ \protect{\hat{\mathcal J}}) \rangle |_{A_{\mathrm{e}}=0} , 
\label{eq:22SR}\\
\Phi_{\mathrm{m}}^{\mathrm{I}} [
\zeta | a_{\mathrm{e}}
 ] 
 & = 
\protect \langle 
T_-\exp (
-i{\zeta}_- \hat J ) T_+\exp (
i{\zeta}_+ \hat J ) \rangle , 
\label{eq:62ZQ}\end{align}%
cf.\ the ``immediate reservation'' at the end of \mbox{Sec.\ \ref{ch:54BP}}. Consequently, rigorous formal definitions of \mbox{$p$} and \mbox{$p^{\mathrm{I}}$} are given by \mbox{Eqs.\ (\ref{eq:41AZ})}, (\ref{eq:22SR}) and (\ref{eq:24ST}), (\ref{eq:62ZQ}), respectively. 
\section{Linear media}%
\label{ch:LM}%
To account for the linear susceptibility of matter in properties of the field, it suffices to replace \mbox{Eq.\ (\ref{eq:62ZQ})} defining functional \mbox{$\Phi _{\mathrm{m}}^{\mathrm{I}}$} by, 
\begin{align} 
\exp(i\zeta \Pi _{\text{R}} a_{\mathrm{e}})\Phi_{\mathrm{m}}^{\mathrm{I}} [
\zeta | a_{\mathrm{e}}
 ] 
 & = 
\protect \langle 
T_-\exp (
-i{\zeta}_- \hat J ) T_+\exp (
i{\zeta}_+ \hat J ) \rangle , 
\label{eq:75UY}\end{align}%
where 
\begin{align} 
\Pi _{\text{R}} (t,t') = \frac{i}{\hbar }\theta(t-t')
\protect \langle [ \hat J(t),\hat J(t') ] \rangle . 
\label{eq:76UZ}\end{align}%
\mbox{Equation (\ref{eq:75UY})} implies substitution (\ref{eq:23SS}). Then, for passive linear media, 
\begin{align} 
\Phi_{\mathrm{m}}^{\mathrm{I}} [
\zeta | a_{\mathrm{e}}
 ] = 1 . 
\label{eq:77VA}\end{align}%
Susceptibility \mbox{$\Pi _{\text{R}} (t,t')$} may be accounted for by including it in the Dyson equation for the retarded propagator, cf.\ 
Appendix D.1.2 in \cite{DirResp}, or by any other method one would care to employ. 

For interacting devices, redefinition (\ref{eq:75UY}) extends to individual devices,
\begin{align} 
\begin{aligned} 
 & \exp(i\zeta \Pi _{\mathrm{R}A}a_{\mathrm{e}})\Phi_{\mathrm{m}A}^{\mathrm{I}} [
\zeta | a_{\mathrm{e}}
 ] 
 \\&\quad = 
\protect \langle 
T_-\exp (
-i{\zeta}_- \hat J_A
 ) T_+\exp (
i{\zeta}_+ \hat J_A 
 ) \rangle , 
 \\& \Pi _{\mathrm{R}A}(t,t') = \frac{i}{\hbar }\theta(t-t')
\protect \langle [ \hat J_A(t),\hat J_A(t') ] \rangle , 
\end{aligned} 
\label{eq:80VD}\end{align}%
and similarly for device $B$. The all-important \mbox{Eq.\ (\ref{eq:63BF})} persists, with susceptibility of matter being a sum of those of devices,
\begin{align} 
\Pi _{\mathrm{R}}(t,t') = 
\Pi _{\mathrm{R}A}(t,t') + 
\Pi _{\mathrm{R}B}(t,t') , 
\label{eq:78VB}\end{align}%
as expected. If a particular device (say, $B$) is a linear passive one, it is fully accounted for in \mbox{$G_{\text{R}} $} and may be excluded from the explicit consideration, 
\begin{align} 
\Phi_{\mathrm{m}B}^{\mathrm{I}} [
\zeta | a_{\mathrm{e}}
 ] = 1 . 
\label{eq:81VE}\end{align}%

Redefinition (\ref{eq:75UY}) is not compulsory, nor is (\ref{eq:80VD}), nor does anything preclude one from applying (\ref{eq:80VD}) selectively, only to certain devices. 
The latter may be a matter of physical clarity. For example, for a coherent linear quantum amplifier, \mbox{Eq.\ (\ref{eq:80VD})} separates its amplification properties (expressed by the imaginary part of \mbox{$\Pi _{\mathrm{R}}$}) from its noise properties (accounted for in \mbox{$\Phi_{\mathrm{m}}^{\mathrm{I}}$}). In quantum mechanics, these two properties are connected \cite{Caves}. Formally separating them is not an error, but may obscure the physical picture.

\section{The two-device problem in the conventional closed-time-loop\ formalism}%
\label{ch:2DC}%
For simplicity we confine the discussion to current. As a starting point we employ Eq.\ (56) of preprint \cite{QDynResp}. With dropped sources, reduced to current and adapted to the present notation it reads, 
\begin{align} 
\Phi_{\mathrm{m}}[{\zeta}_+,{\zeta}_- ] 
& = \exp\bigg(
-\frac{i\hbar}{2}\frac{\delta }{\delta A_+} G _{\mathrm{F}} \frac{\delta }{\delta A_+} 
+\frac{i\hbar}{2}\frac{\delta }{\delta A_-} G_{\mathrm{F}}^* \frac{\delta }{\delta A_-}
\nonumber \\ & \quad 
-i\hbar \frac{\delta }{\delta A_-} G^{(+)} \frac{\delta }{\delta A_+}
\bigg)
\nonumber\\ & \qquad \times 
\Phi_{\mathrm{m}}^{\mathrm{I}} \bigg[
\frac{ A_+ }{\hbar }+\zeta_+,
\frac{ A_- }{\hbar }+\zeta_-
\bigg]\Big|_{ A_{\pm}=0}, 
\label{eq:86AU} %
\end{align}%
where \mbox{$\Phi_{\mathrm{m}} $} and \mbox{$\Phi_{\mathrm{m}}^{\mathrm{I}}$} are defined by \mbox{Eqs.\ (\ref{eq:22SR})} and (\ref{eq:62ZQ}), \mbox{$A_{\mathrm{\pm}}(t)$} are auxilliary functions, and the kernels \mbox{$G_{\text{F}}$} and \mbox{$G^{(+)}$} are the so-called contractions, 
\begin{align} 
\begin{aligned} & G_{\mathrm{F}}(t-t') = \frac{i}{\hbar}\langle 0|
T_+\hat A(t)\hat A(t')
|0\rangle , 
\\
 & G^{(+)}(t-t') = \frac{i}{\hbar}\langle 0|
\hat A(t)\hat A(t')
|0\rangle . \end{aligned} 
\label{eq:87LL}\end{align}%
We use an obvious extension of notation (\ref{eq:76NU}). \mbox{Equation (\ref{eq:86AU})} originates in Wick's theorem. For details see \cite{QDynResp,WickCaus}. 

Adapting manipulations (\ref{eq:95FM})--(\ref{eq:65BJ}) to \mbox{Eq.\ (\ref{eq:86AU})} results in the closed-time-loop\ prototype of \mbox{Eq.\ (\ref{eq:66BK})},
\begin{widetext} 
\begin{align} 
\Phi_{\mathrm{m}}[{\zeta}_+,{\zeta}_- ] 
& = \exp\bigg(
-i\hbar\frac{\delta }{\delta A_+} G _{\mathrm{F}} \frac{\delta }{\delta A_+'} 
+i\hbar\frac{\delta }{\delta A_-} G_{\mathrm{F}}^* \frac{\delta }{\delta A_-'}
-i\hbar \frac{\delta }{\delta A_-} G^{(+)} \frac{\delta }{\delta A_+'}
-i\hbar \frac{\delta }{\delta A_-'} G^{(+)} \frac{\delta }{\delta A_+}
\bigg)
\nonumber\\ & \quad \times 
\Phi_{\mathrm{m}A} \bigg[
\frac{ A_+ }{\hbar }+\zeta_+,
\frac{ A_- }{\hbar }+\zeta_-
\bigg]
\Phi_{\mathrm{m}B} \bigg[
\frac{ A_+ '}{\hbar }+\zeta_+,
\frac{ A_- '}{\hbar }+\zeta_-
\bigg]
\Big|_{ A_{\pm}=A_{\pm}'=0}, 
\label{eq:73UW}\end{align}%
where we also used symmetry of \mbox{$G_{\text{F}}$}. Its structure is illustrated symbolically in \mbox{Fig.\ \ref{fig:BirdsEye10}}a.

To recover (\ref{eq:66BK}), (\ref{eq:23SS}) should be supplemented by two more substitutions \cite{QDynResp}, 
\begin{align} 
\begin{gathered} A_{\pm}(t) = a_{\mathrm{e}}'(t)\pm\hbar \zeta^{\prime(\mp)}(t), \quad
 A'_{\pm}(t) = a_{\mathrm{e}}''(t)\pm\hbar \zeta^{\prime\prime(\mp)}(t). \end{gathered} 
\label{eq:38JH} %
\end{align}%
This turns (\ref{eq:73UW}) into, 
\begin{align} 
\Phi_{\mathrm{m}} [ 
\zeta 
| 
a_{\mathrm{e}} 
 ] & = \exp \bigg ( 
-i\frac{\delta }{\delta a_{\mathrm{e}}''} G_{\text{R}} \frac{\delta }{\delta \zeta '} 
-i\frac{\delta }{\delta a_{\mathrm{e}}'} G_{\text{R}} \frac{\delta }{\delta \zeta'' }
 \bigg ) %
\Phi_{\mathrm{m}A} [ 
\zeta + \zeta '
| 
a_{\mathrm{e}} + a_{\mathrm{e}} '
 ] %
\Phi_{\mathrm{m}B} [ 
\zeta+ \zeta ''
| 
a_{\mathrm{e}} + a_{\mathrm{e}} ''
 ] |_{\zeta '=\zeta''= a_{\mathrm{e}}'=a_{\mathrm{e}}''=0} , 
\label{eq:74UX}\end{align}%
which is equivalent to (\ref{eq:66BK}). 
\section{Generalised photodetection problem}%
\label{ch:PhDet}%

The generalised photodetection problem (\mbox{Fig.\ \ref{fig:BirdsEye11}}a) is the only case in the paper when the single-mode approximation must be lifted. It suffices to assume that the electromagnetic field\ comprises two modes, \mbox{$\hat A_{\mathrm{i}}(t)$} and \mbox{$\hat A_{\mathrm{o}}(t)$}, where i,o refer to the {\em input\/} and {\em output\/} of the photodetector. All other quantities are similarly doubled, \mbox{$\hat J(t)\to\hat J_{\mathrm{i,o}}(t)$}, etc.; \mbox{Eq.\ (\ref{eq:61PX})} applies with \mbox{$k=\text{i,o}$}. We shall also refer to the input and output modes as the optical and photocurrent ones. At this stage, all such terms are conditional.

Adapted to two modes, \mbox{Eq.\ (\ref{eq:8KV})} becomes,
\begin{align} 
p [ 
\mathcal{J}_{\mathrm{i}},\mathcal{J}_{\mathrm{o}} | A_{\mathrm{ei}}, A_{\mathrm{eo}}
 ] & = \int \mathcal{D}[\mathcal{J}_{A\mathrm{i}}] \mathcal{D}[\mathcal{J}_{A\mathrm{o}}] \mathcal{D}[\mathcal{J}_{A\mathrm{i}}] \mathcal{D}[\mathcal{J}_{B\mathrm{o}}] 
\delta [\mathcal{J}_{\mathrm{i}}-\mathcal{J}_{A\mathrm{i}}-\mathcal{J}_{B\mathrm{i}}]
\delta [\mathcal{J}_{\mathrm{o}}-\mathcal{J}_{A\mathrm{o}}-\mathcal{J}_{B\mathrm{o}}]
 \nonumber\\&\quad \times 
p_A[\mathcal{J}_{A\mathrm{i}},\mathcal{J}_{A\mathrm{o}} | A_{\mathrm{ei}}+G_{\mathrm{Ri}}\mathcal{J}_{B\mathrm{i}},A_{\mathrm{eo}}+G_{\mathrm{Ro}}\mathcal{J}_{B\mathrm{o}}] 
p_B[\mathcal{J}_{B\mathrm{i}},\mathcal{J}_{B\mathrm{o}} | A_{\mathrm{ei}}+G_{\mathrm{Ri}}\mathcal{J}_{A\mathrm{i}},A_{\mathrm{eo}}+G_{\mathrm{Ro}}\mathcal{J}_{A\mathrm{o}}] 
, 
\label{eq:60PW}\end{align}%
\end{widetext}%
where, 
\begin{align} 
\begin{aligned} 
G_{\mathrm{R}{\mathrm{i}}}(t-t') & = \frac{i}{\hbar }\theta(t-t') 
 \big [ 
\hat A_{{\mathrm{i}}}(t),\hat A_{{\mathrm{i}}}(t')
 \big ] , \\ 
G_{\mathrm{R}{\mathrm{o}}}(t-t') & = \frac{i}{\hbar }\theta(t-t') 
 \big [ 
\hat A_{{\mathrm{o}}}(t),\hat A_{{\mathrm{o}}}(t')
 \big ] . 
\end{aligned} 
\label{eq:54AT}\end{align}%
The only assumption made when obtaining (\ref{eq:60PW}) is that \mbox{$\hat A_{{\mathrm{i}}}(t)$} and \mbox{$\hat A_{{\mathrm{o}}}(t)$} commute

Equation (\ref{eq:60PW}) is not yet a photodetection formula. It is symmetric with respect to the modes, as well as to the devices; electromagnetic interaction\ in it is bi-directional. It also contains a lot of irrelevant information, in particular, full quantum response properties of the input and output fields. It ``knows'' how an attempt to measure the input field would affect the detection, and how simultaneous measurements of the input field and output current would be correlated (say). Subject to valid quantum models of the devices, it is also ``aware'' of all limitations imposed by quantum mechanics on such simultaneous measurements. 

To suppress the irrelevant information we limit our attention to the quantity, 
\begin{align} 
p [ \mathcal{J}_{\mathrm{o}} ] = \int \mathcal{D}[\mathcal{J}_{\mathrm{i}}]\,p [ 
\mathcal{J}_{\mathrm{i}},\mathcal{J}_{\mathrm{o}} | 0, 0
 ] . 
\label{eq:62PY}\end{align}%
That is, we do not attempt to measure the ``optical'' current, nor to probe the system by radiation in either mode. In fact, it would be easy to preserve \mbox{$A_{\mathrm{ei}}(t)$} as an argument, thus covering photodetection with local oscillators, but we are not currently interested in such generalisation. 

The critical assumption turning device $A$ (say) into a source is that it is not sensitive to the incoming field in either mode, 
\begin{align} 
p_A[\mathcal{J}_{A\mathrm{i}},\mathcal{J}_{A\mathrm{o}} | A_{\mathrm{ei}},A_{\mathrm{eo}}] \approx p_A[\mathcal{J}_{A\mathrm{i}},\mathcal{J}_{A\mathrm{o}} | 0,0]. 
\label{eq:63PZ}\end{align}%
We shall also assume that its contribution to the photocurrent may be neglected, 
\begin{gather} 
p_A[\mathcal{J}_{A\mathrm{i}},\mathcal{J}_{A\mathrm{o}} | 0,0] \approx p_{\mathrm{S}}[\mathcal{J}_{A\mathrm{i}}]\,\delta [ \mathcal{J}_{A\mathrm{o}} ] , 
\label{eq:64QA}\end{gather}%
where
\begin{align} 
p_{\mathrm{S}}[\mathcal{J}_{A\mathrm{i}}] = \int \mathcal{D}[\mathcal{J}_{A\mathrm{o}}]\,p_A[\mathcal{J}_{A\mathrm{i}},\mathcal{J}_{A\mathrm{o}} | 0,0] . 
\label{eq:66QC}\end{align}%
Combining (\ref{eq:60PW}), (\ref{eq:62PY}), (\ref{eq:63PZ}), and (\ref{eq:64QA}) we obtain, 
\begin{align} p [ \mathcal{J}_{\mathrm{o}} ] = \int \mathcal{D}[\mathcal{J}_{A\mathrm{i}}]\,p_{\mathrm{S}}[\mathcal{J}_{A\mathrm{i}}]\,p_{\mathrm{D}}[\mathcal{J}_{\mathrm{o}}|G_{\mathrm{Ri}}\mathcal{J}_{A\mathrm{i}}], 
\label{eq:67QD}\end{align}%
where 
\begin{align} 
p_{\mathrm{D}}[\mathcal{J}_{\mathrm{o}}|A_{\mathrm{i}}] = \int \mathcal{D}[\mathcal{J}_{A\mathrm{i}}]\,p_B[\mathcal{J}_{A\mathrm{i}},\mathcal{J}_{\mathrm{o}} | A_{\mathrm{i}},0] . 
\label{eq:68QE}\end{align}%
Classical collocations of \mbox{Eq.\ (\ref{eq:67QD})} are unmistakeable. It could be guessed from the very beginning. The actual result of this appendix are \mbox{Eqs.\ (\ref{eq:66QC})} and (\ref{eq:68QE}). They define properties of the source and detector in strictly quantum terms, including {\em quantisation of the photocurrent\/}.

To establish a closer connection with \mbox{Sec.\ \ref{ch:61NP}}, we introduce the characteristic functional \mbox{$\Phi _{\mathrm{S}}(\eta )$} of averages (\ref{eq:13SF}). The operator \mbox{$\protect{\hat{\mathcal A}}(t)$} occuring there should be identified with \mbox{$\protect{\hat{\mathcal A}}_{{\mathrm{i}}}(t)$}, defined within the model of solitary device $A$. 
This way,
\begin{align} 
\Phi _{\mathrm{S}}(\eta ) & = \protect \langle {\mathcal T}{\mbox{\rm\boldmath$:$}}\exp
 ( i\eta \protect{\hat{\mathcal A}}_{{\mathrm{i}}}) {\mbox{\rm\boldmath$:$}}\rangle _{A,0} \nonumber\\& = \int \mathcal{D}[\mathcal{J}_{A\mathrm{i}}]\,p_{\mathrm{S}}[\mathcal{J}_{A\mathrm{i}}]\,\exp ( i\eta G_{\mathrm{Ri}}\mathcal{J}_{A\mathrm{i}} ) . 
\label{eq:70QH}\end{align}%
The additional subscript 0 stands for the condition, 
\begin{align} 
A_{\mathrm{ei}}(t) = 
A_{\mathrm{eo}}(t) = 0. 
\label{eq:71QJ}\end{align}%
Functional (\ref{eq:70QH}) is a formal expression of the source$+$field entity, encircled in the dashed oval in \mbox{Fig.\ \ref{fig:BirdsEye11}}a. With it, Eq.\ (\ref{eq:67QD}) may be rewritten as,
\begin{align} 
p [ \mathcal{J}_{\mathrm{o}} ] = \Phi _{\mathrm{S}} \Big ( -i\frac{\delta }{\delta A_{\mathrm{ei}}} \Big ) p_{\mathrm{D}}[\mathcal{J}_{\mathrm{o}}|A_{\mathrm{ei}}]|_{A_{\mathrm{ei}}=0} .
\label{eq:69QF}\end{align}%
That averages (\ref{eq:13SF}) indeed fully characterise the source is evident.


\begin{thebibliography}{68}
\bibitem[{end ({\natexlab{a}})}]{endMeso}
The term ``mesoscopic'' refers to quantum manifestations under
 macroscopic conditions. Common as it actually is, this term is confusing. It
 implies that there exists something ``else'' that distinguishes macroscopic
 and mesoscopic conditions. In fact, there is nothing ``else'' apart from the
 hard-working experimental physicist, rescuing fragile quantum effects from
 being destroyed by environmental and measurement noises. We regard these
 terms as synonyms, with a tendency to apply ``macroscopic'' to experiment and
 ``mesoscopic''---to theory  (as in ``macroscopic experiments are described by
 mesoscopic phenomenologies'').
\bibitem[{end ({\natexlab{e}})}]{endLinBosMeso}
The distinction between the mesoscopic and microscopic
 coordinates is very much blurred for linear bosonic systems. This results, we
 believe, in a lot of confusion, because such systems are commonly used as a
 testing ground for fundamental concepts.
\bibitem[{end ({\natexlab{f}})}]{endKuboSchwinger}
The reader familiar with c-number sources according to Schwinger
 will find the discussion of their relation to Kubo's in \cite{APII,APIII}.
\bibitem[{end ({\natexlab{g}})}]{endOSimpDet}
For pedagogical reasons, we oversimplify the discussion of
 photodetection in section \ref{ch:61NP},  brushing under the carpet a number
 of physically important points. We therefore strongly recommend the reader to
 give proper attention to Appendix \ref{ch:PhDet}.
\end{thebibliography}

\end{document}